\documentclass[showpacs,pra,onecolumn]{revtex4-1}
\usepackage[utf8]{inputenc}
\usepackage{amssymb}
\usepackage{epstopdf}
\usepackage{hyperref}
\usepackage{graphicx}
\usepackage{amsmath}
\usepackage{subfigure}
\usepackage{grffile}
\usepackage[sort&compress]{natbib}
\usepackage{color}

\begin{document}

\title{Spontaneous symmetry breaking of self-trapped and leaky modes in
quasi-double-well potentials}
\author{Krzysztof B. Zegadlo$^{1}$, Nir Dror$^{2}$, Marek Trippenbach$^{3}$,
Miroslaw A. Karpierz$^{1}$ and Boris A. Malomed$^{2}$}
\affiliation{$^{1}$Faculty of Physics, Warsaw University of Technology, Warsaw, ul.
Koszykowa 75, PL-00-662 Warszawa, Poland\\
$^{2}$Department of Physical Electronics, School of Electrical Engineering,
Faculty of Engineering, Tel Aviv University, Tel Aviv 69978, Israel\\
$^{3}$Faculty of Physics, University of Warsaw, ul. Ho\.{z}a 69, PL-00-681,
Warszawa, Poland}

\begin{abstract}
We investigate competition between two phase transitions of the second kind
induced by the self-attractive nonlinearity, \textit{viz.}, self-trapping of
the leaky modes, and spontaneous symmetry breaking (SSB) of both fully
trapped and leaky states. We use a one-dimensional mean-field model, which
combines the cubic nonlinearity and a double-well-potential (DWP) structure
with an elevated floor, which supports leaky modes (quasi-bound states) in
the linear limit. The setting can be implemented in nonlinear optics and
BEC. The order in which the SSB and self-trapping transitions take place
with the growth of the nonlinearity strength depends on the height of the
central barrier of the DWP: the SSB happens first if the barrier is
relatively high, while self-trapping comes first if the barrier is lower.
The SSB of the leaky modes is characterized by specific asymmetry of their
radiation tails, which, in addition, feature a resonant dependence on the
relation between the total size of the system and radiation wavelength. As a
result of the SSB, the instability of symmetric modes initiates spontaneous
Josephson oscillations. Collisions of freely moving solitons with the DWP
structure admit trapping of an incident soliton into a state of persistent
shuttle motion, due to emission of radiation. The study is carried out
numerically, and basic results are explained by means of analytical
considerations.
\end{abstract}

\maketitle

\section{Introduction}

Usually, the ground state (GS) of quantum-mechanical systems exactly follows
the symmetry of the underlying Hamiltonian \cite{LL}, while excited states
may realize different representations of the same symmetry (a different
situation is exemplified by the Jahn-Teller effect in molecules, which makes
the GS of the electron subsystem spatially asymmetric, thus breaking the
symmetry of the respective Hamiltonian \cite{RE}). In particular, for the
double-well potential (DWP), which is dealt with in the present work, the GS
wave function is spatially even, while the first excited state is odd. This
is not necessarily true in many-body settings. In that context, the
mean-field description of atomic Bose-Einstein condensates (BECs) is
provided by the Gross-Pitaevskii equation (GPE) \cite{BEC}, which includes
the cubic term accounting for attractive forces between colliding atoms.
Essentially the same is the nonlinear Schr\"{o}dinger equation (NLSE)
modeling the propagation of optical signals in Kerr-nonlinear media \cite%
{NLS}. If the self-focusing nonlinearity is strong enough, it gives rise to
the phase transition in the form of \textit{spontaneous symmetry breaking}
(SSB) of the GS \cite{book}. In its simplest manifestation, which is
provided by the DWP, the SSB implies that one well traps a larger atomic
density or field power than the other. This effect also implies the breakup
of the basic principle of quantum mechanics, according to which the GS
cannot be degenerate, as the SSB gives rise to a pair of two mutually
symmetric GSs in the DWP, with the maximum of the wave function found in
either potential well (as mentioned above, the Jahn-Teller effect gives rise
to a qualitatively similar situation). The same DWP setting admits a
symmetric state coexisting with the asymmetric ones, but, above the SSB
point, the symmetric wave function no longer represents the GS, being
unstable against symmetry-breaking perturbations. In the course of the
spontaneous transition from the unstable symmetric state to a stable
asymmetric one, the choice between the two mutually degenerate asymmetric
states is determined by random perturbations, which push the system to build
the maximum of the wave function in the left or right potential well. The
SSB is a ubiquitous phenomenon, with well-known manifestations in nonlinear
optics, BEC, superfluidity, superconductivity, ferromagnetism, etc. \cite%
{book}.

The concept of the SSB in nonlinear systems of the NLS type was, plausibly,
introduced for the first time in 1979 by E. B. Davies \cite{Davies}, who
addressed a nonlinear extension of the Schr\"{o}dinger equation for a pair
of quantum particles with an isotropic interaction potential. In this
context, the SSB was predicted as the breaking of the rotational symmetry in
the GS. Another early work, which predicted the SSB in a relatively simple
form, addressed the \textit{self-trapping model}, based on a system of
linearly coupled ordinary differential equations including the
self-attractive cubic terms \cite{Scott}.

In the effectively one-dimensional geometry, the SSB can be studied in the
framework of the scaled NLSE/GPE with potential $H(x)$ of the DWP type, for
the amplitude of the electromagnetic wave, or the single-particle wave
function, $\psi \left( x,z\right) $:
\begin{equation}
i\frac{\partial \psi }{\partial z}=-\frac{1}{2}\frac{\partial ^{2}\psi }{%
\partial x^{2}}-\left\vert \psi \right\vert ^{2}\psi +H(x)\psi ,
\label{NLSE}
\end{equation}%
where $z$ is the propagation distance in optics, or time in the GPE. This
equation can be reduced to a system of coupled ordinary differential
equations for two amplitudes, $u_{1,2}(z)$, by means of the \textit{%
tight-binding approximation} \cite{tight}, which replaces $\psi (x,z)$ by a
linear superposition of two stationary wave functions, $\phi $,
corresponding to the states trapped separately in either potential well,
with their centers located at $x=\pm a$ \cite{Ananikian}:%
\begin{equation}
\psi \left( x,z\right) =u_{1}(z)\phi \left( x-a\right) +u_{2}(z)\phi \left(
x+a\right) .  \label{12}
\end{equation}

The analysis of the SSB in BEC and similar models based on Eq. (\ref{NLSE})
was initiated in Refs. \cite{Milburn} and \cite{Smerzi}. In this case, in
the framework of the mean-field approximation, the symmetry breaking is the
phase transition of the second kind (alias the supercritical bifurcation,
which does not admit hysteresis \cite{Joseph}). Further, GPE (\ref{NLSE})
was extended by adding an extra (free) spatial coordinate, which transforms
the DWP into a two-dimensional dual-core structure \cite{Warsaw}. In such a
setting, the self-attractive nonlinearity gives rise to matter-wave
solitons, which self-trap in the free direction \cite{soliton}. The SSB
destabilizes symmetric solitons and replaces them by asymmetric ones,
provided that the norm of the wave function (which determines the effective
strength of the intrinsic nonlinearity) exceeds a critical value \cite%
{Warsaw}. In the latter case, the mean-field symmetry breaking is a phase
transition of the first kind (alias a subcritical bifurcation \cite{Joseph}%
), which includes hysteresis. The subcritical transition is typical to
solitons in dual-core waveguides with the Kerr self-focusing \cite%
{soliton-dual-core}. The same type of the transition may be featured by CW
(continuous-wave) states in dual-core systems with non-Kerr nonlinearities
\cite{Snyder}.

In addition to the analysis of static symmetric and asymmetric modes,
dynamical regimes, most typically in the form of oscillations of the
mean-field wave function between two wells of the DWP structure, were
analyzed too. Following the analogy with Josephson oscillations of the wave
function of Cooper-paired electrons in superconducting tunnel junctions \cite%
{Ustinov}, the possibility of oscillations in \textit{bosonic Josephson
junctions} was predicted \cite{junction}. The simplest dynamical model of
the Josephson oscillations in bosonic systems was derived by means of the
tight-binding approximation (\cite{superconductor}).

Experimental manifestations of the SSB have been observed in both BEC and
photonics. Self-trapping of a macroscopically asymmetric state of the atomic
condensate of $^{87}$Rb atoms, loaded into the DWP, as well as Josephson
oscillations in that setting, were reported in Ref. \cite{Markus} (in that
case, the effective nonlinearity is self-repulsive, therefore the respective
SSB occurs not in the symmetric GS, but rather in the antisymmetric first
excited state). The SSB of laser beams coupled into an effective transverse
DWP created in the self-focusing photorefractive medium was demonstrated in
Ref. \cite{photo}. Other experimentally observed SSB effects in optics are
spontaneously established asymmetric regimes of operation of coupled lasers
\cite{lasers,NatPhot,NaturePhot2,Chili}, and breaking of chiral symmetry in
metamaterials \cite{Kivshar}.

In addition to usual bound states, one may work with quasi-localized modes
in potentials which do not admit complete trapping in linear quantum
mechanics, but give rise to leaky bound states, alias quasi-bound ones. The
combination of such a potential and self-attractive nonlinearity makes it
possible to transform the leaky states into truly bound ones \cite%
{Carr1,Carr2}. This possibility, in turn, suggests another setting, which is
the subject of the present work: DWP structures embedded into a potential
barrier. In the linear limit, this structure support solely symmetric leaky
modes, that may be transformed into self-trapped ones with the help of the
cubic self-attraction. The main feature of the system which, to the best of
our knowledge, was not explored before, is competition between two different
mean-field phase transitions of the second kind, driven by the nonlinearity:
the SSB and transition to the self-trapping. Realization of the competition
in stationary states of the DWP system is the main subject of the present
work. We demonstrate that, depending on parameters of the DWP structure and
nonlinearity strength, either transition may happen first, with the growth
of the nonlinearity. Another essential problem addressed in the paper is a
dynamical one, namely, Josephson oscillations in the DWP structure,
initiated by the instability of the symmetric mode, and collisions of free
solitons with the structure.

It is relevant to mention that, in terms of the BEC realizations, the
present setting represents macroscopic quantum states, with the phase
transitions between them being quasi-classical ones, considered in the
framework of the mean-field approximation. The validity of this
approximation is usually justified by the large number of atoms in the
condensate \cite{BEC}. The consideration of a few-body state in the DWP can
give rise to quantum phase transitions, such as those in the
Lipkin-Meshkov-Glick model, which applies in this case \cite{LMG}. As
suggested by a recent analysis of the three-dimensional many-body quantum
gas with repulsive binary interactions, which is pulled to the center by a
potential $\sim -1/r^{2}$ \cite{Grisha}, the quantum phase transition may
produce results similar to but different from their mean-field counterparts
\cite{HS}. In particular, the GS predicted by the mean-field may be replaced
by a metastable state in the quantum many-body theory \cite{Grisha}. In any
case, the consideration of truly quantum phase transitions in the DWP
structure is a subject for a separate work.

The subsequent presentation is structured as follows. The model is
elaborated in Section II. Results of the analysis of symmetric and
spontaneously emerging asymmetric trapped and leaky modes in it are
summarized in Section III. Detailed results are obtained in a numerical
form, and their basic features are explained by means of an analytical
approach. Both the trapped and leaky modes undergo the SSB transition with
the increase of the norm, the symmetric modes getting unstable above the
transition point. The nonlinear evolution of the unstable modes, which
features Josephson oscillations, is studied by means of systematic
simulations in Section IV. A related possibility is capture of incident
solitons by the DWP structure into shuttle states. This possibility is
studied in a systematic form in Section V. The paper is concluded by Section
VI.

\section{The model}

The underlying dynamical model, based on Eq. (\ref{NLSE}), gives rise to
stationary modes with propagation constant $k$ (in BEC, $-k$ is the chemical
potential), \ \
\begin{equation}
\psi \left( x,z\right) =e^{ikz}u\left( x\right) ,  \label{psi}
\end{equation}%
where real modal functions $u(x)$ satisfy the equation

\begin{equation}
-ku+\frac{1}{2}u^{\prime \prime }+u^{3}=H\left( x\right) u.
\label{stationaryNLSE}
\end{equation}%
Solutions with $k>0$ represent self-trapped localized states, while $k<0$
corresponds to delocalized leaky modes, which do not vanish at $x\rightarrow
\pm \infty $.\ The states of these two types are characterized,
respectively, by convergent and divergent norms, $N=\int_{-\infty }^{+\infty
}u^{2}(x)dx$ (proportional to the total power of the light beam in optics,
or the total number of atoms in BEC), and Hamiltonian (energy),%
\begin{equation}
E=\int_{-\infty }^{+\infty }\left[ \frac{1}{2}\left( u^{\prime }\right) ^{2}-%
\frac{1}{2}u^{4}+2H(x)u^{2}\right] dx.  \label{H}
\end{equation}

The DWP can be readily implemented in the experiment. In optics, waveguides
with this structure are fabricated with the help of the implanting technique
\cite{Opt-DWP}, while in BEC the DWP setting can be created by means of
electromagnetic fields \cite{BEC-DWP}. In the present work, calculations are
reported for the rectangular DWP profile with the elevated floor:

\begin{equation}
H\left( x\right) =\left\{
\begin{array}{ll}
A, & ~\mathrm{at}~~|x|<0.5, \\
2, & ~\mathrm{at}~~3<|x|<7, \\
0, & ~\mathrm{elsewhere},%
\end{array}%
\right.   \label{potential}
\end{equation}%
where $A>0$ is the height of the inner potential barrier, see Fig. \ref%
{potential-fig}, while height $H_{\max }=2$ of the outer barriers is fixed
by scaling. Values of lengths adopted in Eq. (\ref{potential}) adequately
represent the generic situation, as demonstrated by additional numerical
results (not shown here in detail). Indeed, it is shown below that the
symmetry-breaking and self-trapping transitions, and the competition between
them, crucially depend on the tunneling transparency of the central barrier
and the nonlinearity strength, i.e., the barrier height, $A$, and total
norm, $N$. These are two control parameters which are subject to the
variation in the subsequent analysis. For the same reason, the main findings
are not sensitive to a particular shape of the DWP. In particular, the
rectangular form of the DWP, adopted in Eq. (\ref{potential}), which may be
essential for some dynamical effects, such as temporal scaling in the
relaxation of perturbations \cite{Azbel}, produces results which are
essentially the same as generated by smooth DWP profiles (for the
self-trapping of the leaky modes in a single potential well, this property
was known before \cite{Carr1}). As concerns the necessity of having the
elevated potential floor, it may make the experimental creation of the
structure easier, as the \textquotedblleft floor" is naturally built by
overlap of fringes of two potential barriers which determine the DWP (in
previously reported experimental realizations of the DWP in BEC \cite{Markus}%
, the bottom level of the potential had to be depressed, because those DWPs
trapped condensates with the repulsive interactions, on the contrary to the
case of the self-attraction, considered herein).

\begin{figure}[th]
\centering
\includegraphics[scale=0.3]{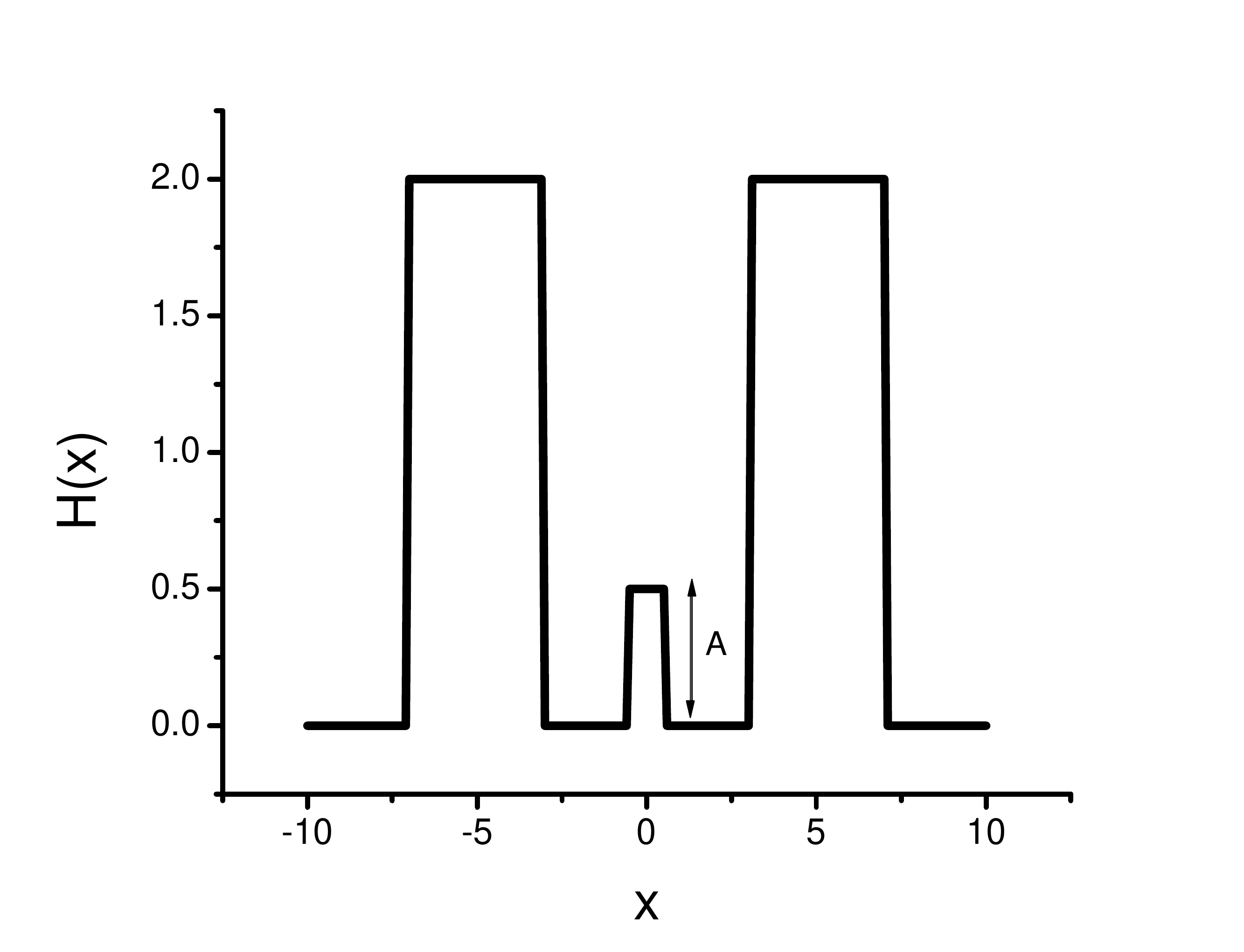}
\caption{The symmetric double-well potential for leaky modes (quasi-bound
states), defined as per Eq. (\protect\ref{potential}).}
\label{potential-fig}
\end{figure}

Obviously, in the absence of the self-attractive cubic term, potential (\ref%
{potential}) cannot support any bound state in the respective linear model,
while weakly delocalized quasi-bound states are possible. Indeed, a
straightforward estimate of the tunneling coefficient for the tall barriers
separating the inner and outer parts of structure (\ref{potential}) yields
\begin{equation}
T\simeq \exp \left( -\sqrt{2H_{\max }-q^{2}}W\right) \approx \allowbreak
4.4\times 10^{-4},  \label{T}
\end{equation}%
where $H_{\max }=2$ and $W=4$ are the height and width of the potential
barriers, as per Eq. (\ref{potential}), and $q\equiv \pi /l=\pi /6$ is the
wavenumber of the lowest quasi-bound state in the potential box of width $%
l=6 $.

For the study of collisions of moving solitons with the DWP structure, the
height of the inner barrier is fixed to be $A=0.5$, while the height, $H_{0}$%
, and width, $W_{0}$, of the outer barriers will be varied, to allow clearer
observations of different collision scenarios:
\begin{equation}
H\left( x\right) =\left\{
\begin{array}{ll}
0.5, & ~\mathrm{at}~~|x|<0.5, \\
H_{0}, & ~\mathrm{at}~~3<|x|<3+W_{0}\equiv \Lambda /2, \\
0, & ~\mathrm{elsewhere}.%
\end{array}%
\right.  \label{P2}
\end{equation}

Detailed consideration of the SSB in the leaky modes will require an
explicit calculation of small-amplitude radiation tails attached to those
modes outside of the barriers, i.e., at $|x|>7$, see Eq. (\ref{potential}).
For this purpose, the DWP is embedded into a broad free-space domain, $%
|x|<L/2$, with zero boundary conditions (b.c.):
\begin{equation}
u(|x|=L/2)=0.  \label{bc}
\end{equation}

\section{Spontaneous symmetry breaking (SSB) of leaky and trapped modes}

\subsection{The structure of symmetric and asymmetric modes}

Numerical solutions of Eq. (\ref{stationaryNLSE}) were obtained by means of
the shooting and Newton-matrix methods. While the former one makes it
possible to find all relevant solutions independently of an input trial
function, the latter method can be applied to obtain solution with high
accuracy, provided that the initial guess is taken not too far from the
final result. For the setting addressed in this paper, the combination of
both algorithms is the most efficient way of obtaining stationary solutions.

As is typical for the SSB in systems with self-focusing nonlinearity, it was
found that the GS is spatially symmetric below the bifurcation point ($k<k_{%
\mathrm{bif}}$) and asymmetric above it, at $k>k_{\mathrm{bif}}$. The
symmetric state exists at $k>k_{\mathrm{bif}}$ too, but in that case it is
not a GS, and is no longer stable. As mentioned above, all mean-field phase
transitions exhibited by the present system are of the second kind,
featuring no hysteresis or bistability between symmetric and asymmetric
modes.

\begin{figure}[th]
\centering\subfigure[]{\includegraphics[scale=0.32]{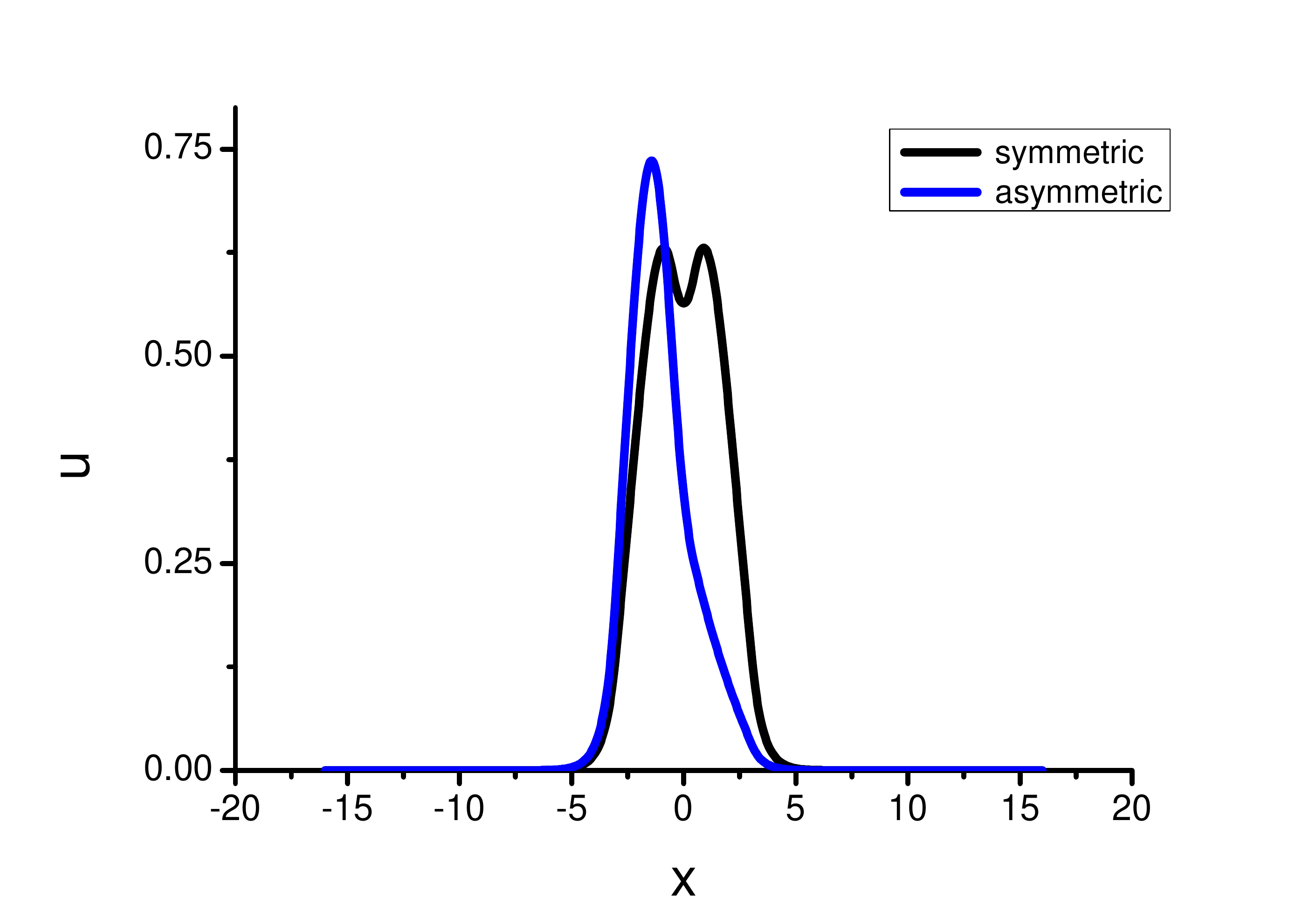}%
\label{trapped_modes}}~\subfigure[]{%
\includegraphics[scale=0.32]{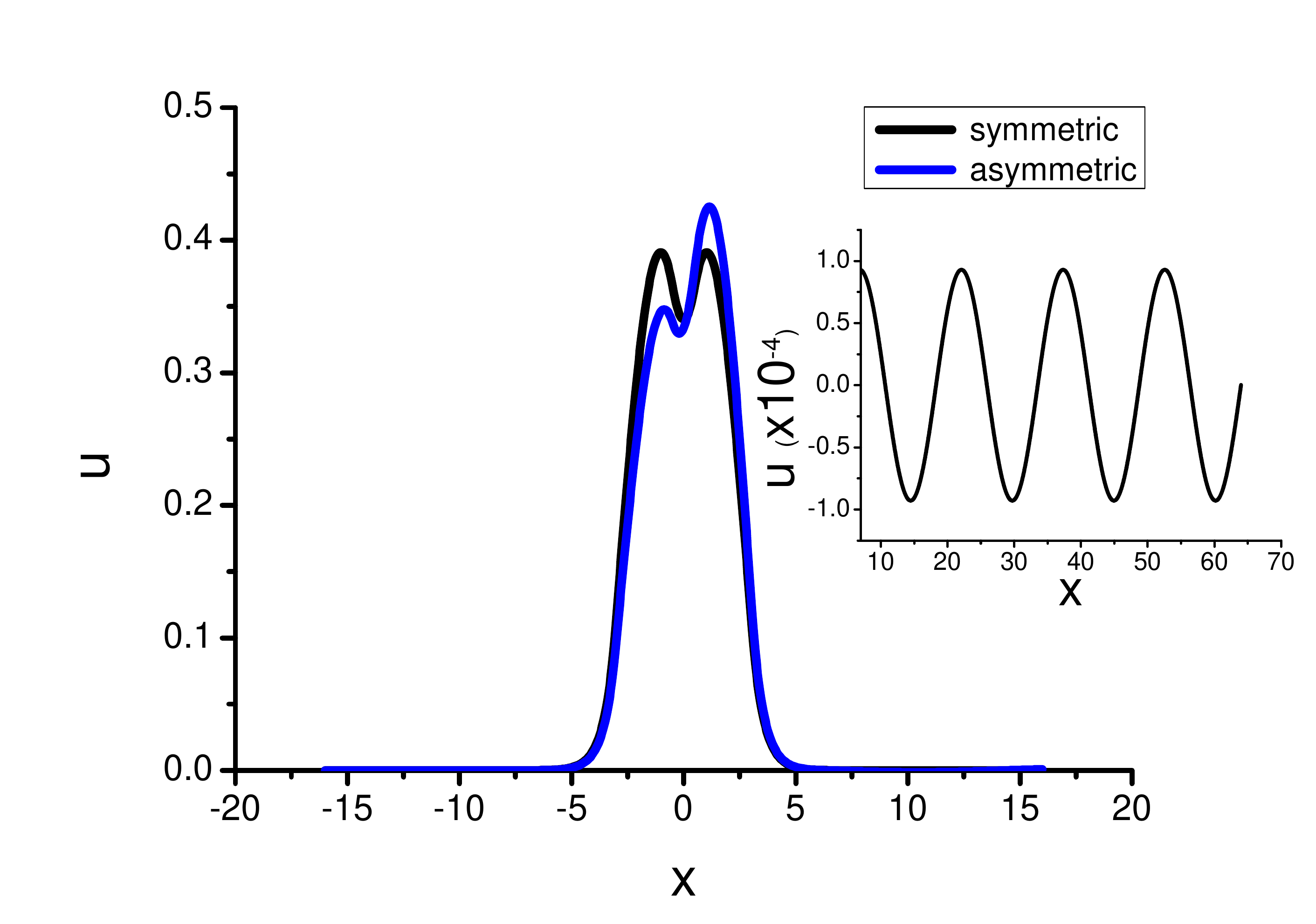}\label{radiated_modes}} %
\subfigure[]{\includegraphics[scale=0.25]{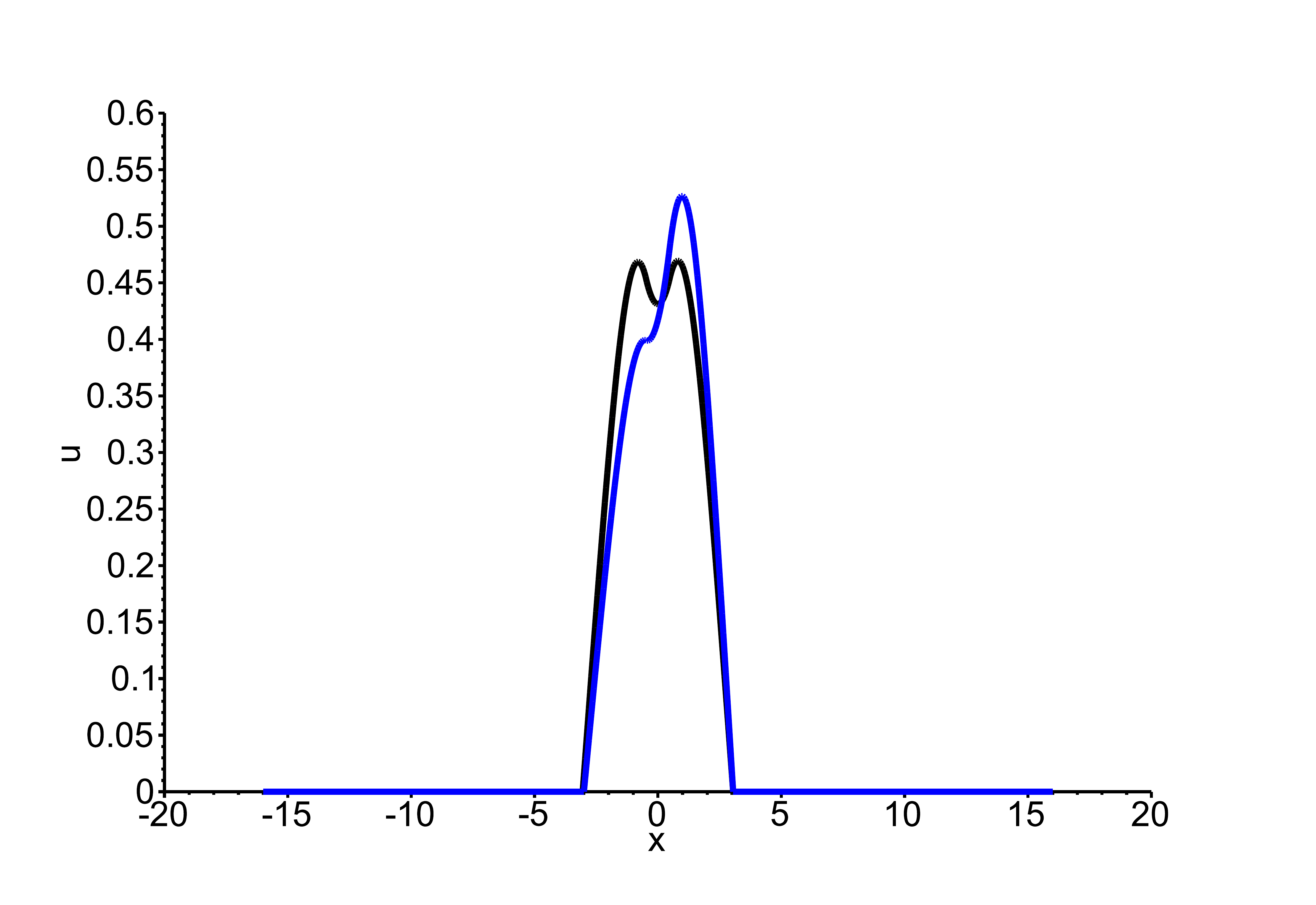}%
\label{radiated_modes}}
\caption{(Color online) (a) Typical examples of trapped symmetric (black)
and asymmetric (blue, gray in printed version) states for propagation constant $k=0.1$, the
respective norms being $N_{\mathrm{symm}}=1.542$ and $N_{\mathrm{asym}}=1.082
$. (b) The same for leaky modes, at $k=-0.085$, with $N_{\mathrm{symm}}=0.642
$ and $N_{\mathrm{asym}}=0.630$. The inset in (b) zooms in on one spatial
period the small oscillatory tail of the symmetric leaky mode. For the sake
of comparison, panel (c) displays the symmetric and asymmetric states with
the same $k$ as in (b), but trapped in DWP (\protect\ref{potential}) with
impenetrable outer barriers (i.e., $H=2$ is replaced by $H=\infty $ in
them), the respective norms being $N_{\mathrm{symm}}=0.792$ and $N_{\mathrm{%
asym}}=0.769$. All the modes are produced by potential (\protect\ref%
{potential}) with the height of the inner barrier $A=0.5$.}
\label{symmetric and asymmetric modes}
\end{figure}

Generic examples of unstable symmetric and stable asymmetric states of both
trapped ($k>0$) the leaky ($k_{\mathrm{bif}}<k<0$) types are displayed in
Figs. \ref{symmetric and asymmetric modes}(a) and (b), respectively. In the
latter case, the delocalized tails of the leaky mode are extremely small,
with amplitude
\begin{equation}
u_{\mathrm{rad}}^{(\max )}\approx 1.2\times 10^{-4}.  \label{rad}
\end{equation}%
Taking into regard the value of the amplitude of the delocalized mode at its
center, $U_{0}\approx 0.40$, tunneling coefficient (\ref{T}) predicts
amplitude $u_{\mathrm{rad}}^{(\max )}\sim TU_{0}\simeq 1.8\times 10^{-4}$,
in reasonable agreement with its numerically found counterpart (\ref{rad}).
For given $k$, the spatial period of the tail in the free space is expected
to be $\lambda =\pi \sqrt{2/|k|}\approx \allowbreak 15$ for $k=-0.085$ in
Fig. \ref{symmetric and asymmetric modes}(b), while the numerical solutions
demonstrates a close value, $\lambda \approx 13.5$ (a small deviation from
the predicted value may be explained by the proximity of the tail to the
outer barrier).

The formally diverging contribution of the tails to the total norm of the
leaky mode is negligible, $N_{\mathrm{rad}}\simeq \left( L/2\right) \left[
u_{\mathrm{rad}}^{(\max )}\right] ^{2}$, where $L\approx 20$ is the total
size of the free-space part of the configuration displayed in Fig. \ref%
{symmetric and asymmetric modes}(b). Indeed, using estimate (\ref{rad}) for
the tail's amplitude yields $N_{\mathrm{rad}}\sim 10^{-7}$, therefore the
leaky modes have definite values of the norm, as indicated in the caption to
Fig. \ref{symmetric and asymmetric modes}(b).

For the sake of direct comparison between leaky and trapped modes, in Fig. %
\ref{symmetric and asymmetric modes}(c) we display stationary states with
the same propagation constant as in Fig. \ref{symmetric and asymmetric modes}%
(b), but in the case when they are confined by impenetrable (infinitely
tall) outer potential barriers in Eq. (\ref{potential}). It is seen that,
similar to the situation shown in Fig. \ref{symmetric and asymmetric modes}%
(b), the smaller and larger values of $N$ correspond to the broken
asymmetric and unbroken symmetric modes, respectively. However, in the
infinitely deep potential box the profiles of the wave functions are,
naturally, narrower and taller.

The symmetric modes displayed in Fig. \ref{symmetric and asymmetric modes}
feature split peaks, due to the fact that the inner potential barrier in Eq.
(\ref{potential}) is relatively high. On the other hand, the same potential
structure with an essentially smaller barrier's height $A$ supports
single-peak symmetric modes, see Fig. \ref{unstable evolution} below.

\subsection{SSB of radiation tails in leaky modes}

As mentioned above, the SSB of trapped modes is a known effect, which was
previously studied in other forms \cite{book,NaturePhot2}. A new phenomenon
reported here is the SSB of leaky modes, which include nonvanishing tails
extending into the free space outside of the DWP structure. Even if the
tails have small amplitudes, it is interesting to analyze their structure in
asymmetric modes, as this issue was not considered previously. To this end,
we here focus on the symmetric and asymmetric states in the setting based on
the DWP (\ref{potential}) with $A=0.5$ and $\Lambda =14$, embedded into a
broad domain of size $L=128$, see Eqs. (\ref{potential}),\ (\ref{P2}), and (%
\ref{bc}). In this case, the asymmetric modes are found at $k\geq -0.100$.

\begin{figure}[th]
\centering
\subfigure[]{\includegraphics[scale=0.3]{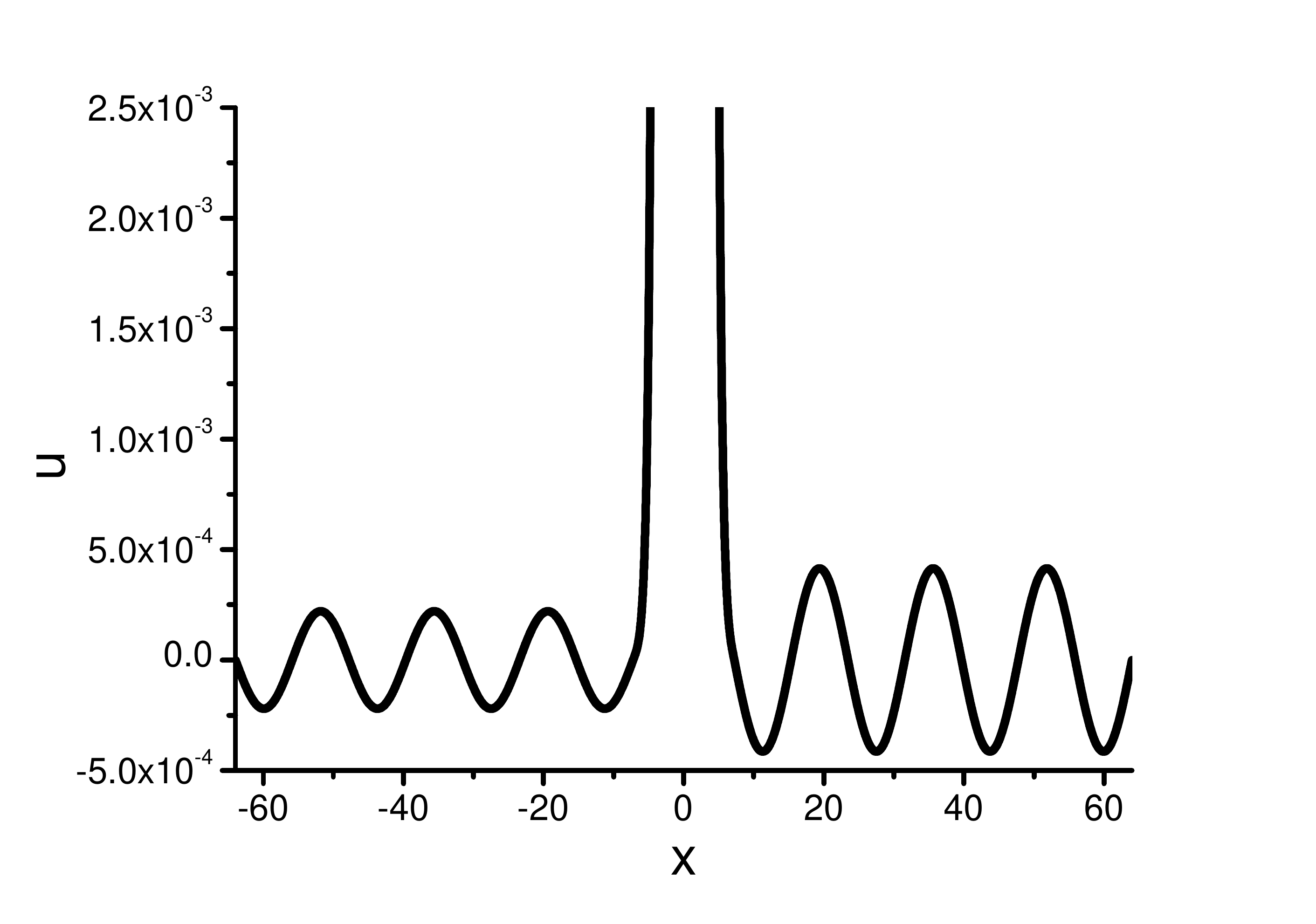}\label{k=-0.075}} %
\subfigure[]{\includegraphics[scale=0.3]{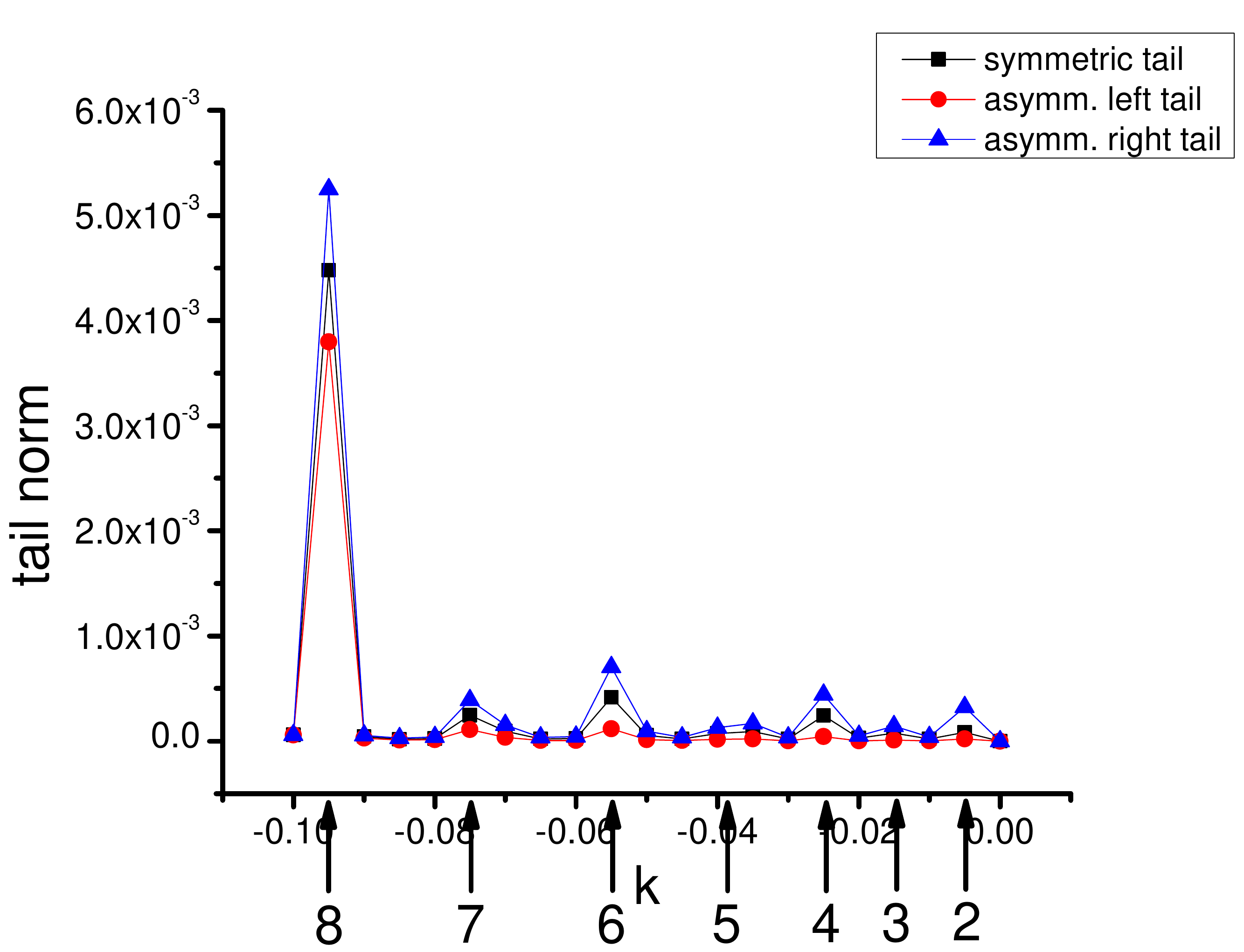}} %
\subfigure[]{\includegraphics[scale=0.3]{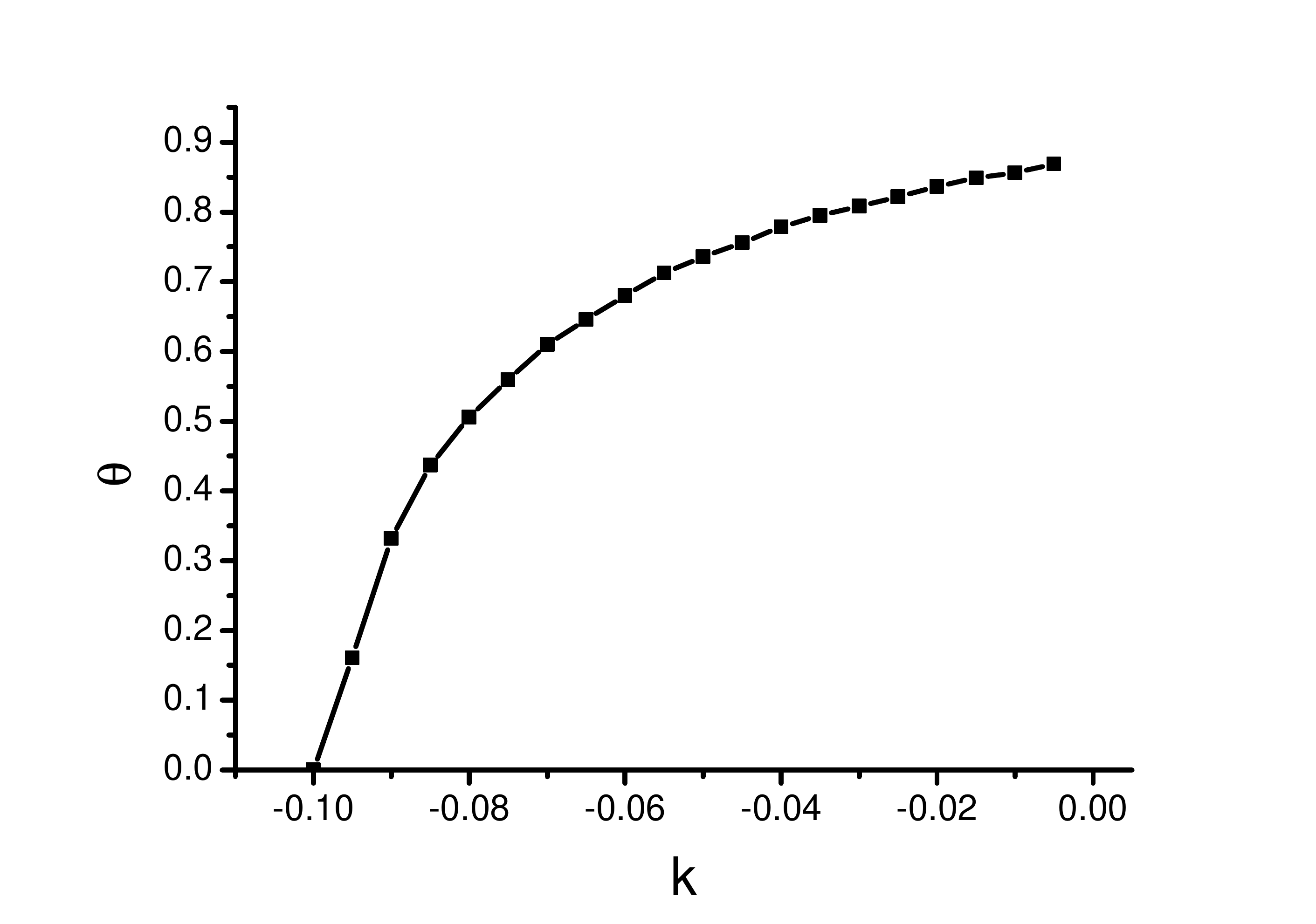}}
\caption{(Color online) (a) A typical example of tails of an asymmetric
leaky mode, found at propagation constant $k=-0.075$. (b) The total norms of
the right and left tails of the asymmetric leaky mode, along with the tail
norm for its symmetric counterpart, vs. $k$. Vertical arrows indicate
positions of the maxima predicted, through the commensurability condition,
by Eq. (\protect\ref{k}). (c). The asymmetry parameter for the tails' norm,
defined as per Eq. (\protect\ref{theta_tail}), vs. $k$. These results were
obtained for $A=0.5$ in DWP structure (\protect\ref{potential}), and the
total size of the system $L=128$.}
\label{tails}
\end{figure}

A characteristic example of \textit{asymmetric tails}, i.e., left and right
ones with unequal amplitudes, is shown in Fig. \ref{tails}(a) for $k=-0.075$%
. Further, separately calculated total norms of the right and left tails, in
regions $\Lambda /2<x<L/2$ and $-\Lambda /2<x<-L/2$, along with the norm of
the tails in the coexisting unstable symmetric leaky mode, are displayed, as
functions of $k$, in Fig. \ref{tails}(b). This dependence exhibits two
notable features. First, the asymmetry between the right and left tails
emerges at $k=-0.100$ and gradually increases with the increase of $k$
(i.e., decrease of $|k|$), even if each tail's norm vanishes in the limit of
$k$\-$\rightarrow 0$ (when the transition to the self-trapped mode takes
place, and the tails vanish). This feature is illustrated by Fig. \ref{tails}%
(c), which displays the asymmetry measure vs. $k$:%
\begin{equation}
\theta (k)\equiv \frac{\int_{\Lambda /2}^{L/2}u^{2}(x;k)dx-\int_{-\Lambda
/2}^{-L/2}u^{2}(x;k)dx}{\int_{\Lambda /2}^{L/2}u^{2}(x;k)dx+\int_{-\Lambda
/2}^{-L/2}u^{2}(x;k)dx}~.  \label{theta_tail}
\end{equation}%
Second, the dependence of the tails' norms on $k$ shows strong oscillations,
which is explained by the commensurability-incommensurability transitions
between the wavelength of the radiation tail and the total size of the
free-space domains, $L/2-\Lambda /2$. Indeed, the radiation wavenumber given
by the free-space dispersion relation for linearized equation (\ref%
{stationaryNLSE}), $q=\sqrt{-2k}$, determines the the radiation
half-wavelength, $\pi /q$, which, in the case of the commensurability,
satisfies relation $(\pi /q)n=L/2-\Lambda /2$, with $n=1,2,3,...$ . Thus,
maxima of the radiation amplitude are expected at discrete values of the
propagation constant, \ \
\begin{equation}
k_{n}=-2\left[ \pi n/\left( L-\Lambda \right) \right] ^{2}.  \label{k}
\end{equation}%
As shown in Fig. \ref{tails}(b), Eq. (\ref{k}) quite accurately predicts
positions of the tail-norm peaks for $n=2,3,4,5,6,7,$ and $8$, for $%
L-\Lambda =114$, which corresponds to the present case [$n=1$ yields $%
k_{1}\approx -1.5\times 10^{-3}$, for which the tail's amplitudes are too
small to discern the corresponding maximum, while Eq. (\ref{k}) with $n=9$
predicts $k_{9}\approx -0.123$, for which asymmetric modes do not exist).

\subsection{SSB diagrams}

\begin{figure}[th]
\centering
\subfigure[] {\includegraphics[scale=0.2]{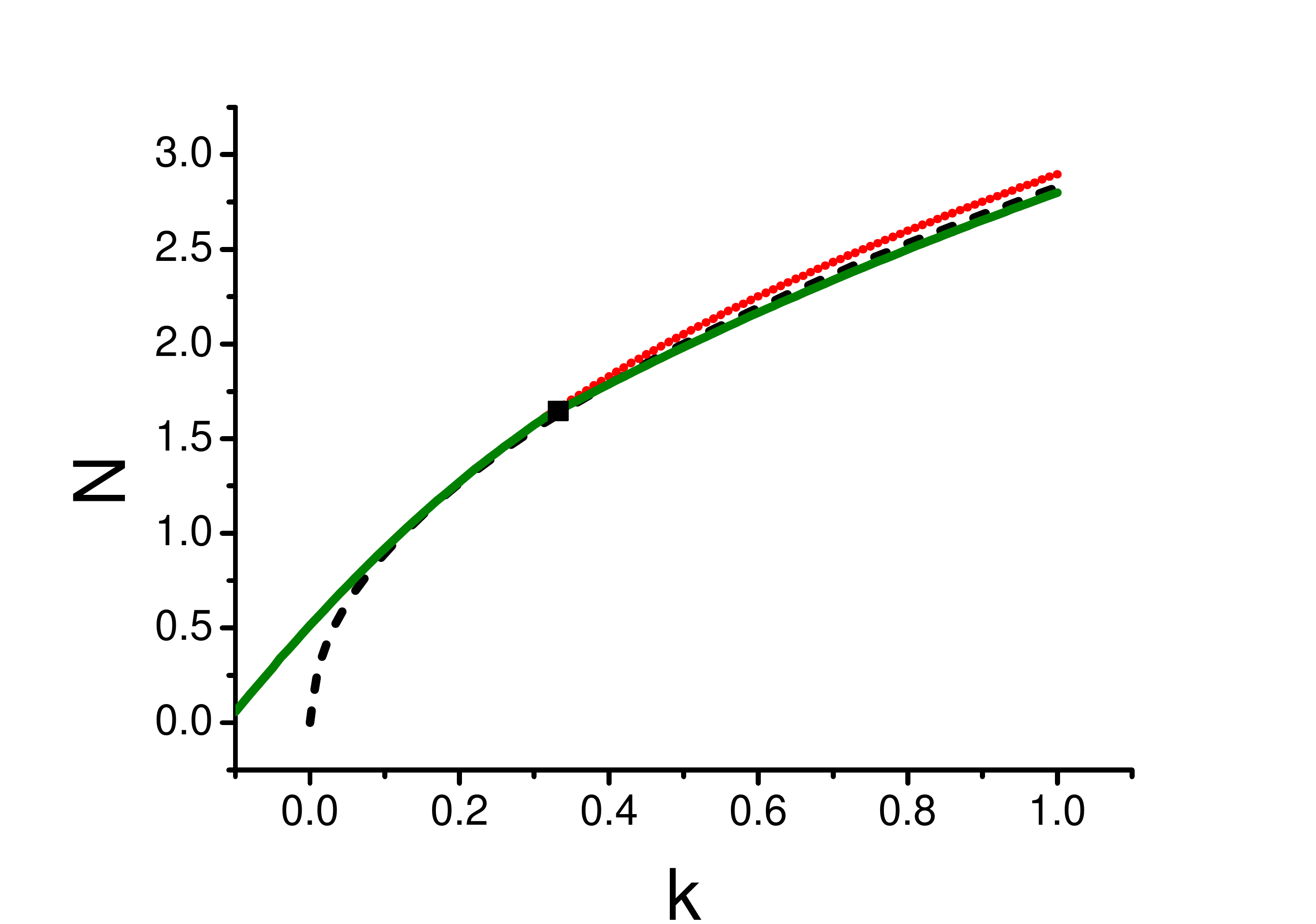} \label{N(k)1}}%
\nolinebreak
\subfigure[] {\includegraphics[scale=0.2]{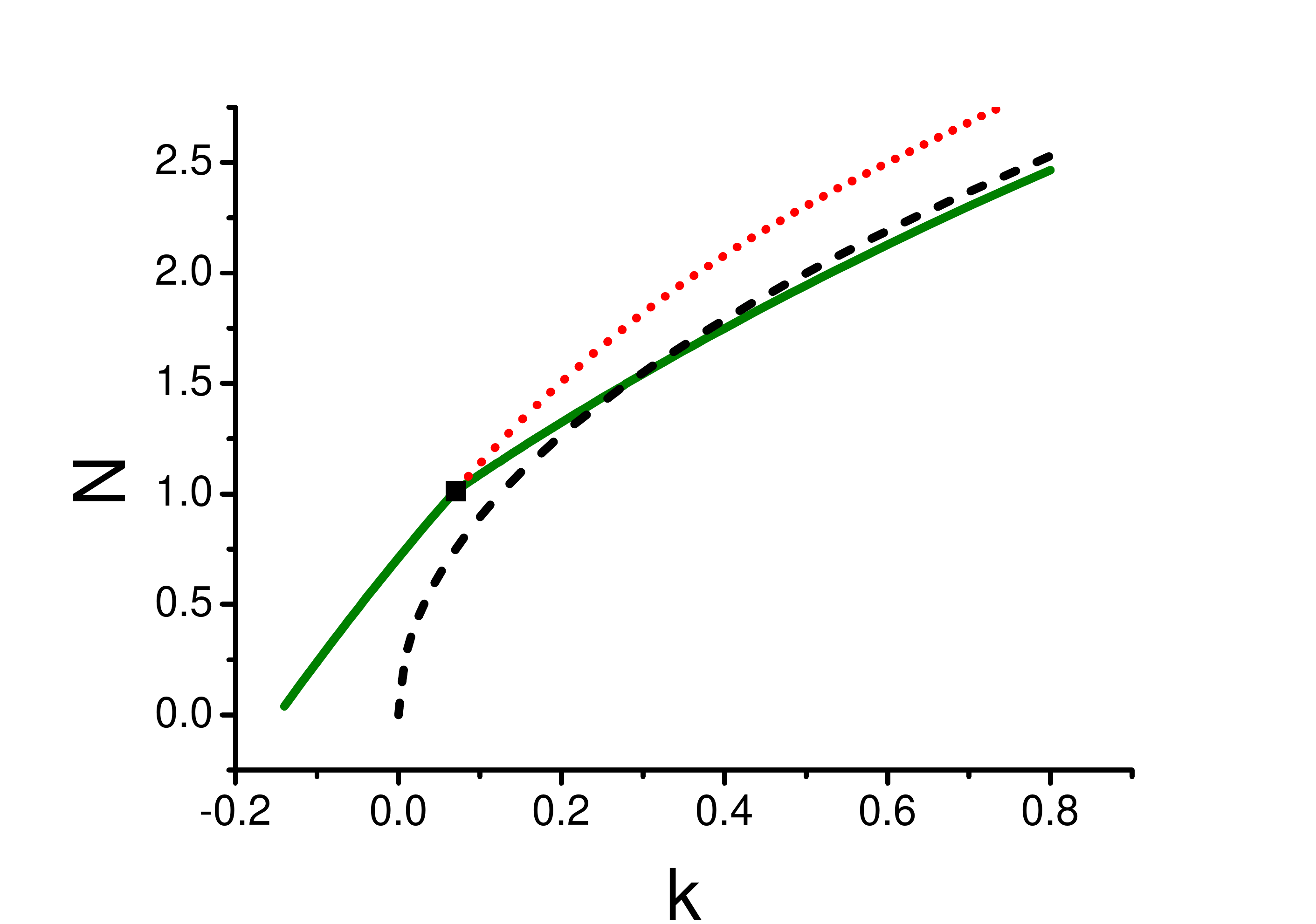}
\label{N(k)2}}\nolinebreak \subfigure[]
{\includegraphics[scale=0.2]{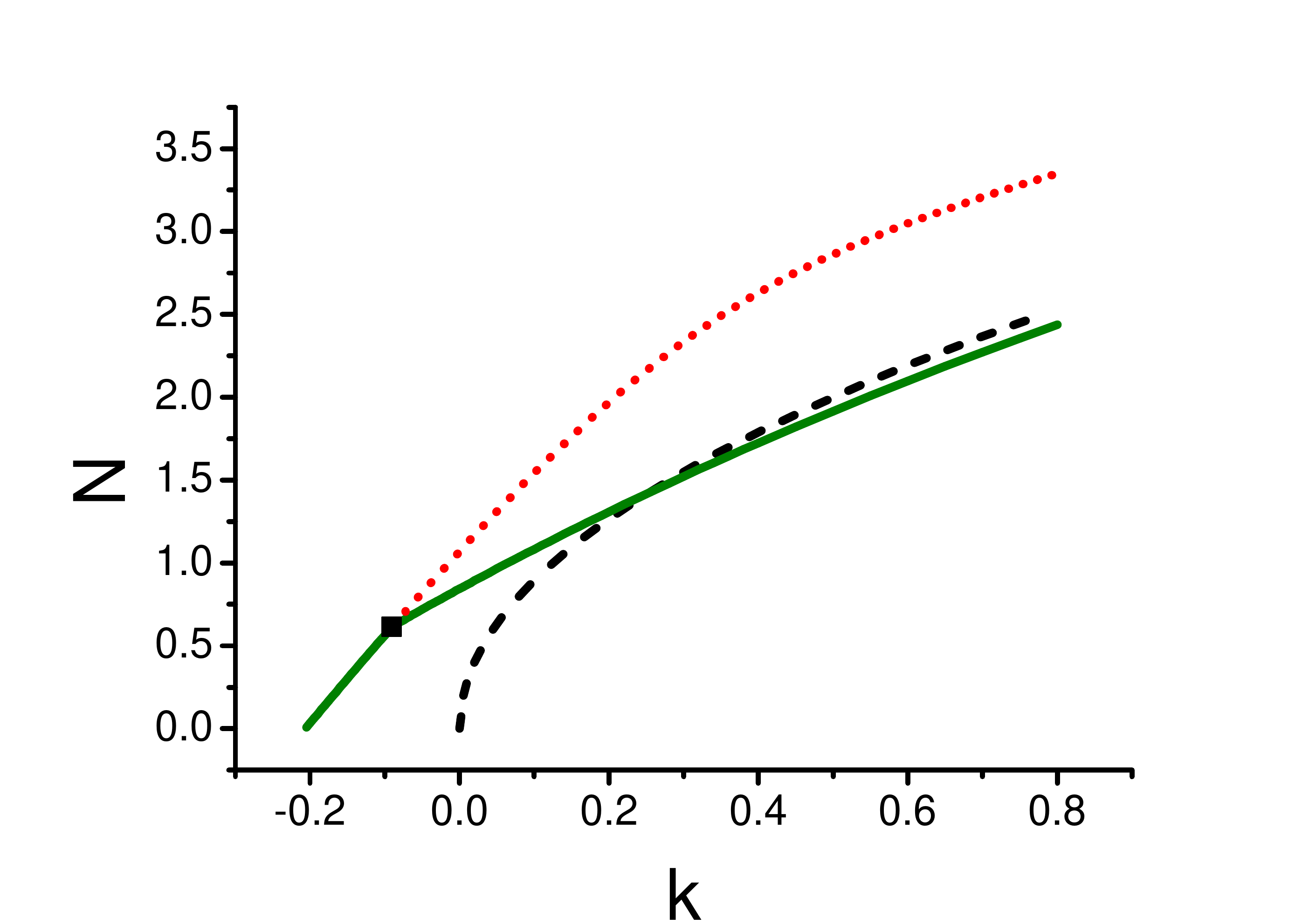} \label{N(k)3}} \subfigure[]
{\includegraphics[scale=0.2]{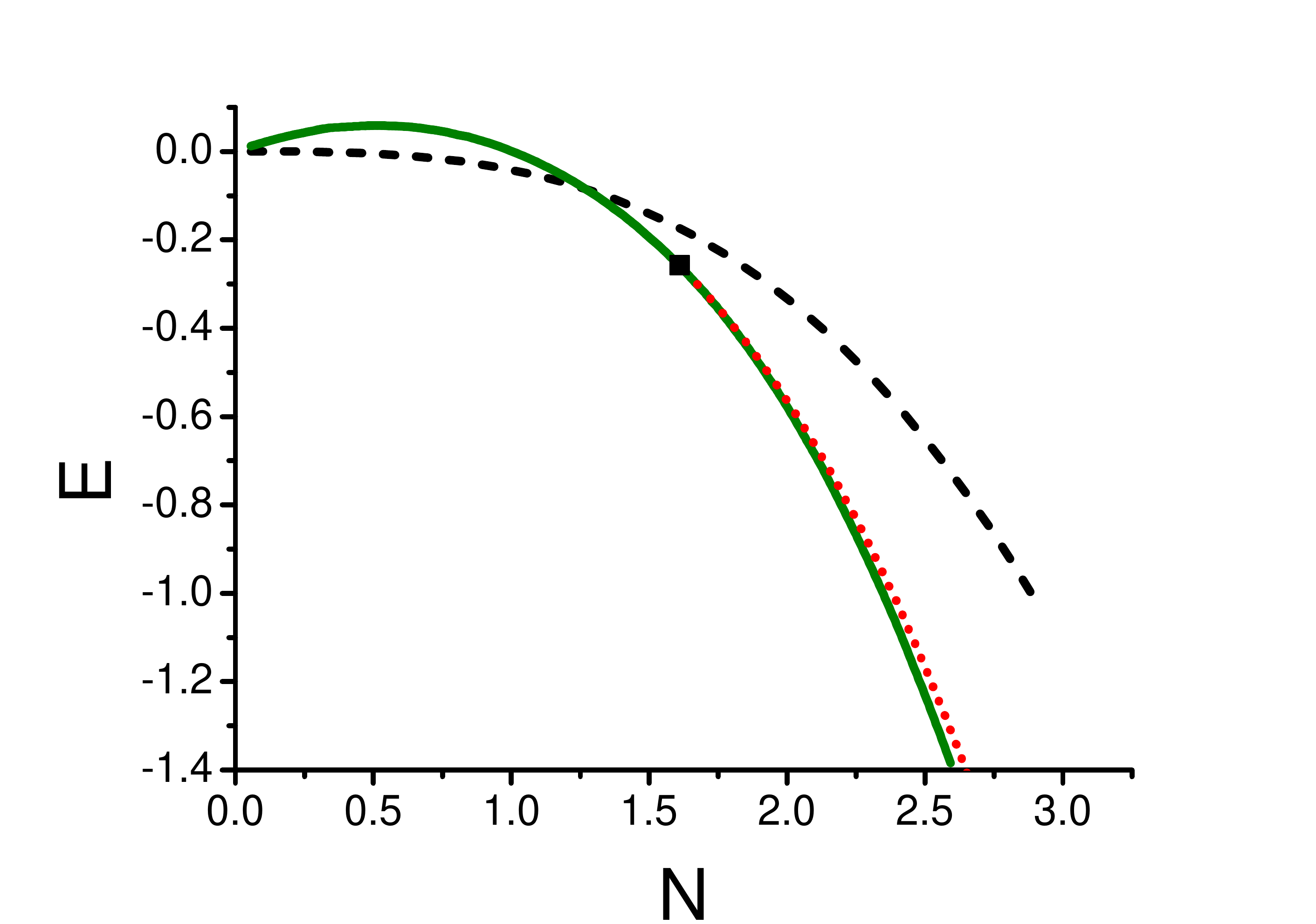} \label{E(N)1}}\nolinebreak %
\subfigure[]{\includegraphics[scale=0.2]{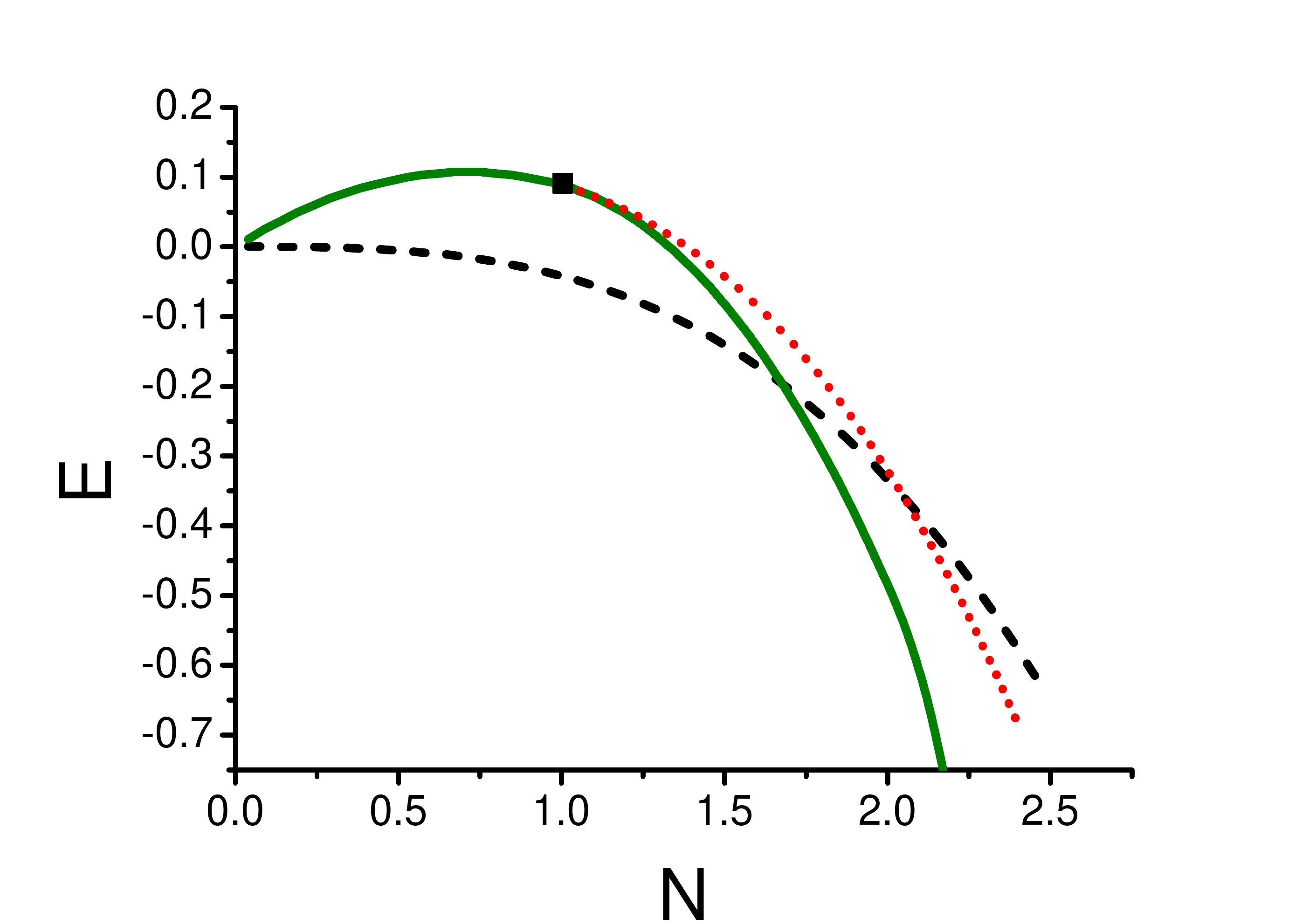} \label{E(N)2}}
\nolinebreak
\subfigure[]{\includegraphics[scale=0.2]{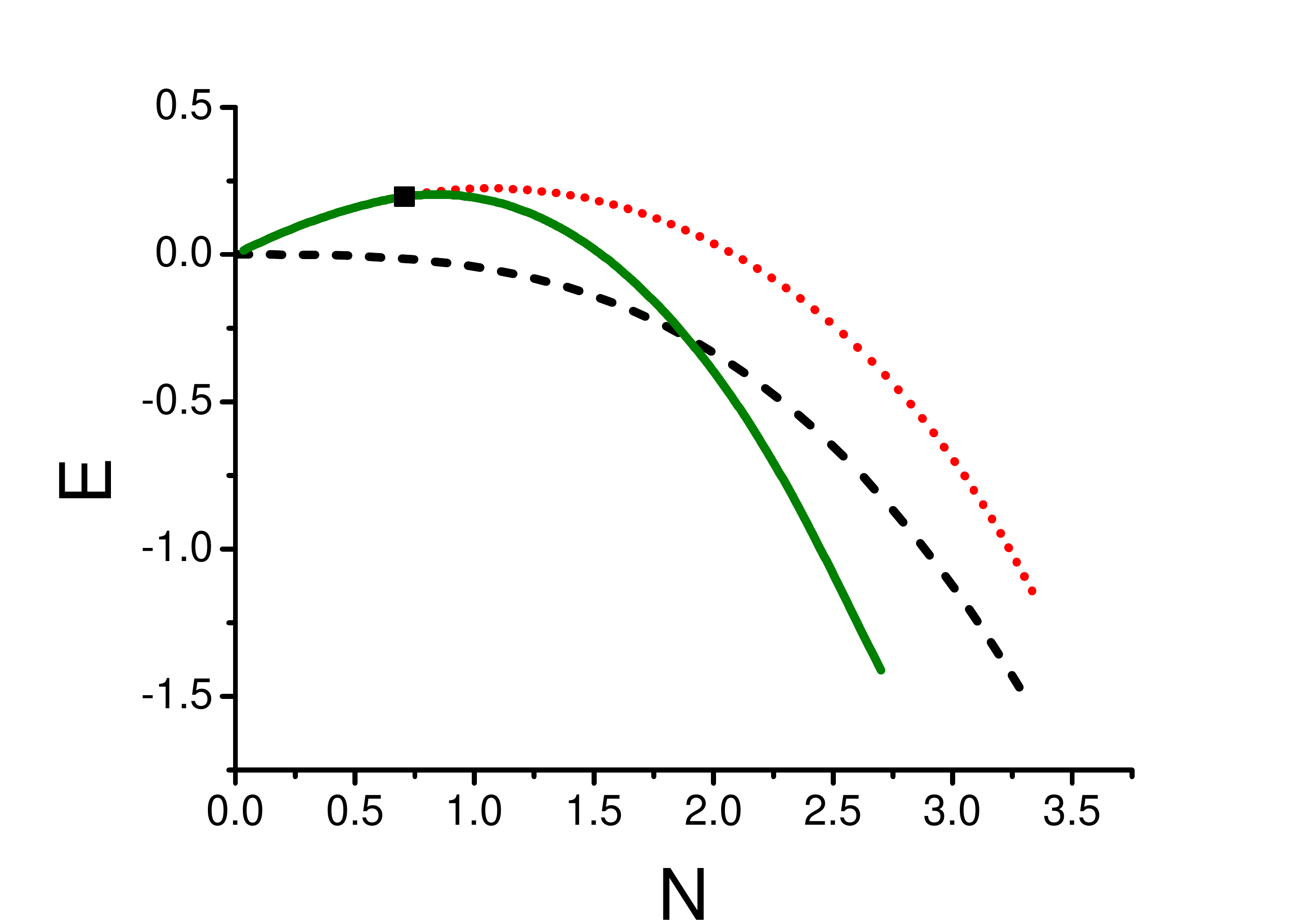}
\label{E(N)3}}\newline
\subfigure[]{\includegraphics[scale=0.2]{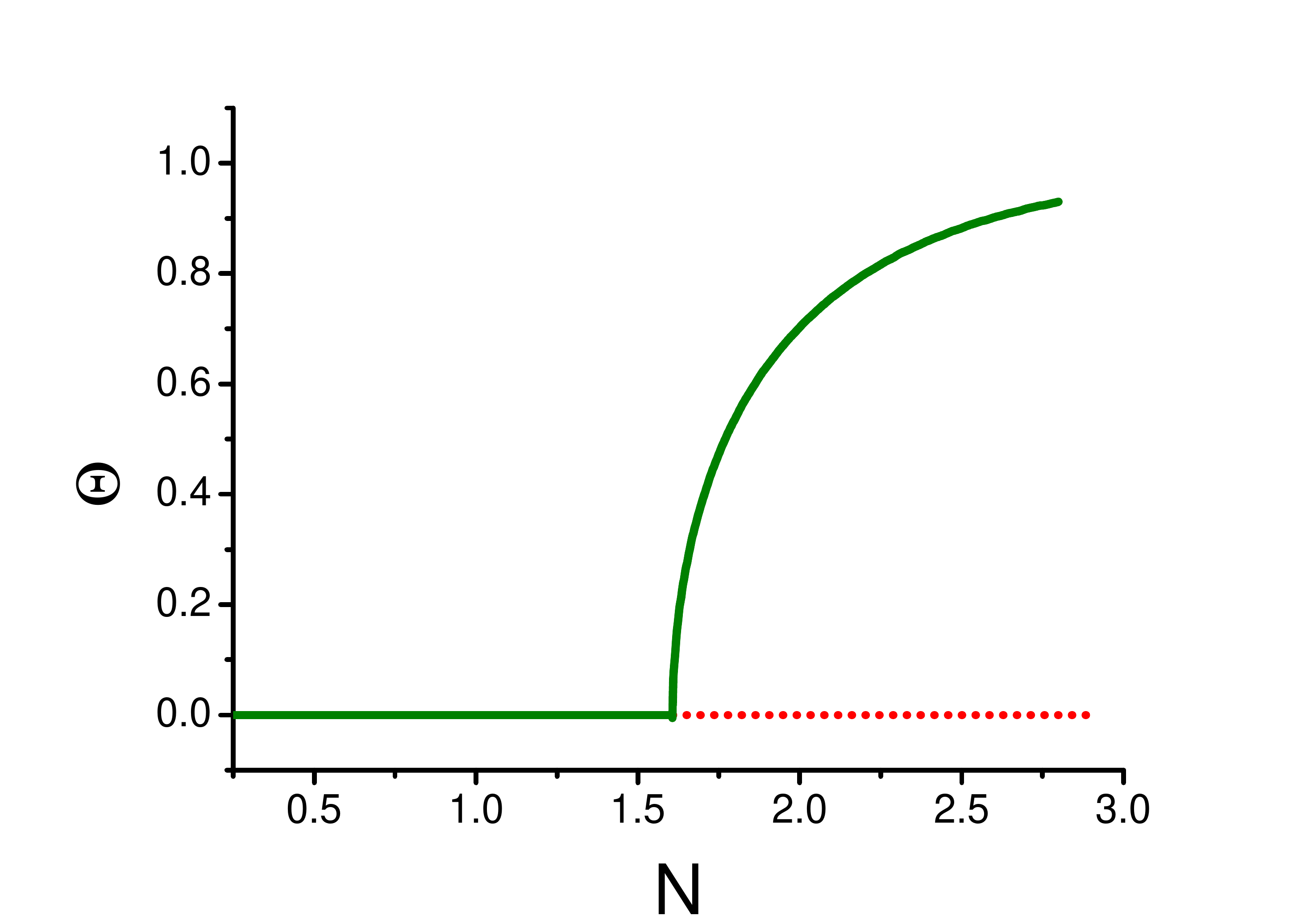} \label{Theta(N)1}}%
\nolinebreak
\subfigure[] {\includegraphics[scale=0.2]{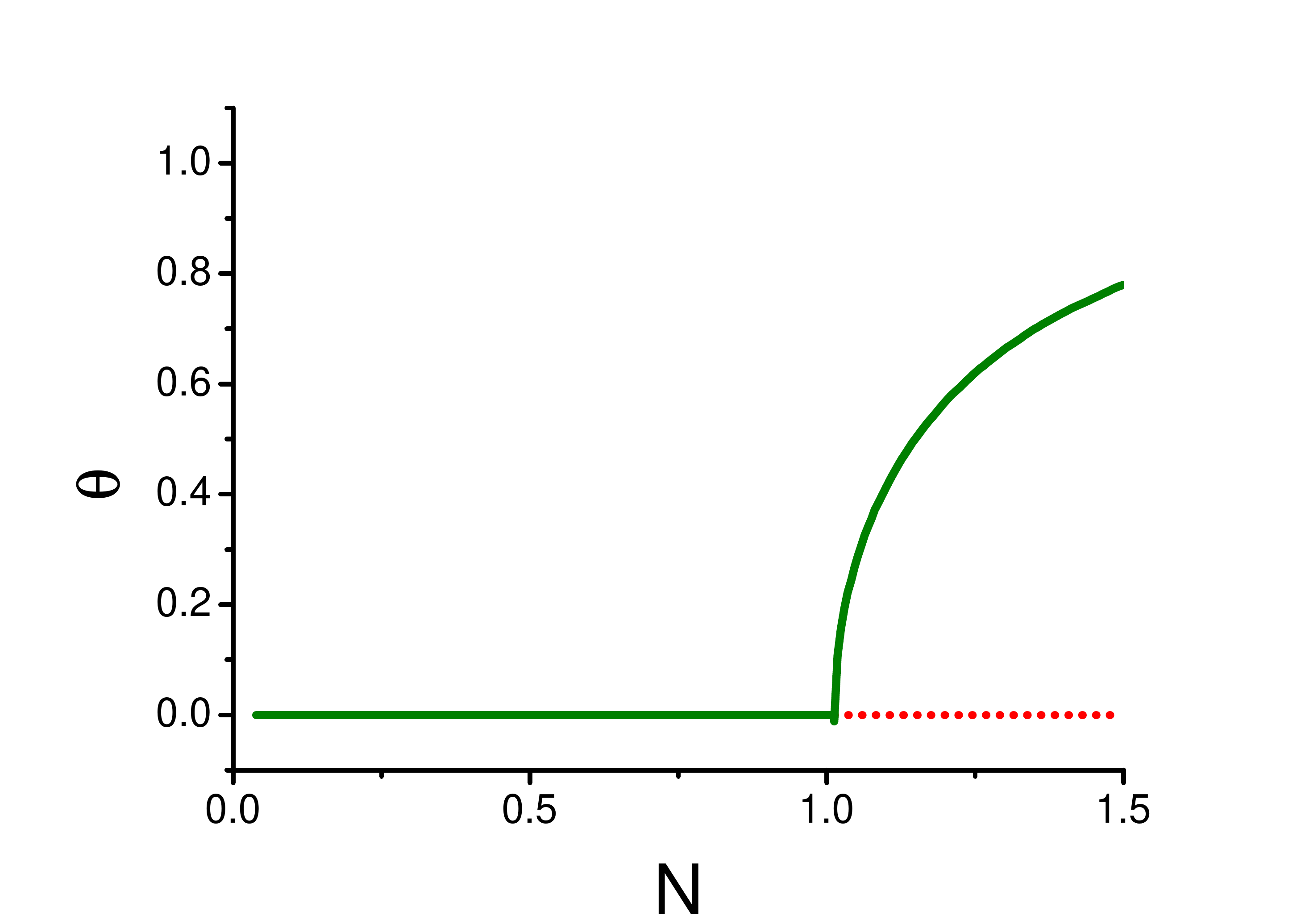}
\label{Theta(N)2}}\nolinebreak
\subfigure[] {\includegraphics[scale=0.2]{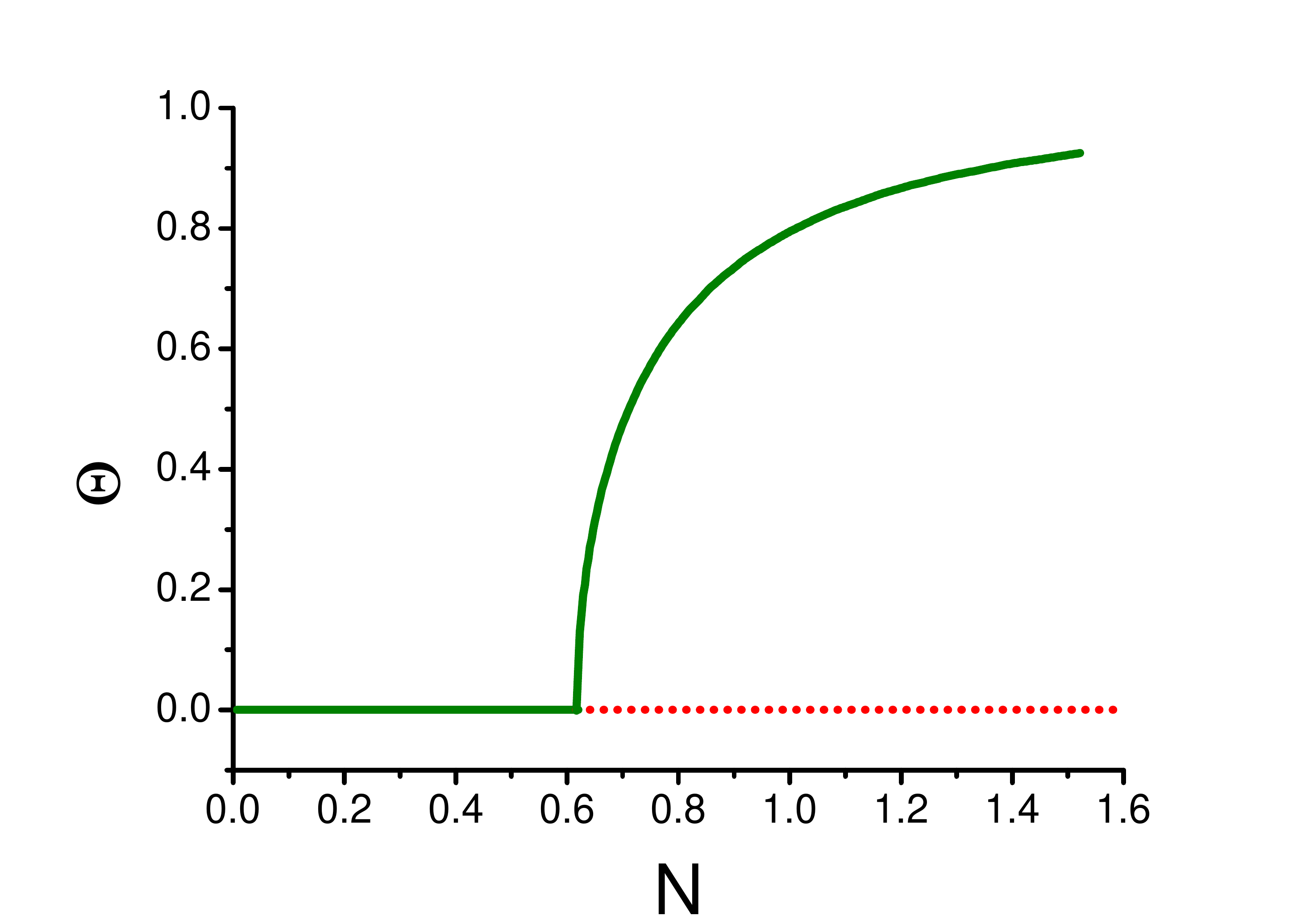}
\label{Theta(N)3}}
\caption{(Color online) Families of symmetric and asymmetric modes,
represented by curves for the norm vs. the propagation constant in panels
(a-c), Hamiltonian vs. the norm in panels (d-f), and the asymmetry [defined
per Eq. (\protect\ref{theta})] vs. the norm in panels (g-i). The left,
middle, and right plots correspond, severally, to the height of the central
barrier in potential (\protect\ref{potential}) $A=0.05$, $0.2$, and $0.5$.
Continuous (olive) and dotted (red) lines designate stable and unstable
states, respectively. The dashed lines in the top and middle panels depict,
for the sake of comparison, the $N(k)$ and $E(N)$ dependences for free-space
solitons, given by Eq. (\protect\ref{sol}). The square symbols in (a-f)
designate the location of the bifurcation points.}
\label{bifurcation diagrams}
\end{figure}

Getting back to the consideration of the SSB for the entire system,
systematic results are presented by means of plots $N(k)$ and $E(N)$ [the
Hamiltonian is defined by Eq. (\ref{H})] for symmetric and asymmetric modes,
which are displayed in Figs. \ref{bifurcation diagrams}(a-c) and (d-f) for
three different values of height $A$ of the inner barrier of potential
structure (\ref{potential}). Note that the $N(k)$ curves obey the
Vakhitov-Kolokolov criterion \cite{VK,VK2}, which is necessary but not
sufficient for the stability of modes supported by the self-attractive
nonlinearity (it does not catch the instability of the symmetric modes
coexisting with asymmetric ones). It is relevant to compare these plots with
their counterparts,%
\begin{equation}
N_{\mathrm{sol}}=2\sqrt{2k},~E_{\mathrm{sol}}=-(1/3)\left( 2k\right) ^{3/2}%
\text{,}  \label{sol}
\end{equation}%
for the NLS solitons in the free space, given by Eq. (\ref{psi}) with%
\begin{equation}
u_{\mathrm{sol}}(x)=\sqrt{2k}\,\mathrm{sech}\left( \sqrt{2k}x\right) ,
\label{soliton}
\end{equation}%
which are displayed by dashed curves in Figs. \ref{bifurcation diagrams}%
(a-f).

The SSB in the families of trapped and leaky states is characterized by the
asymmetry ratio,

\begin{equation}
\Theta \equiv N^{-1}\left[ \int_{0}^{+\infty }u^{2}(x)dx-\int_{-\infty
}^{0}u^{2}(x)dx\right]  \label{theta}
\end{equation}%
[cf. a similar definition for the tails, given by Eq. (\ref{theta_tail})],
which is shown as a function of $N$ in Figs. \ref{bifurcation diagrams}%
(g-i). These plots clearly identify the SSB-onset points, at which the
symmetric mode gets destabilized, and simultaneously a stable asymmetric one
emerges. In accordance with what is said above, the bifurcation is of the
supercritical type \cite{Joseph}, i.e., the emerging branches of the
asymmetric states immediately go \textquotedblleft forward". Conclusions
about the stability and instability of the solution branches displayed in
Fig. \ref{bifurcation diagrams} were produced by means of the well-known
method \cite{VK2} based on numerical computation of (in)stability
eigenvalues (imaginary parts of eigenfrequencies) for small perturbations
added to the stationary solutions, using the respective linearized
(Bogoliubov - de Gennes) equations. In particular, the instability of those
symmetric states which coexist with asymmetric ones is always represented by
a single pair of purely imaginary eigenfrequencies.

The fact that the asymmetric modes, when they exists, have smaller values of
$E$ for given $N$, hence they realize the system's GS [see Figs. \ref%
{bifurcation diagrams}(d-f)], can be easily understood: having their center
shifted from the layer occupied by the positive potential ($|x|<0.5$)\ to a
region where $H=0$ [$0.5<|x|<3$, see Figs. \ref{potential-fig} and \ref%
{symmetric and asymmetric modes}], they obviously reduce the integral value
of $E$. The same argument explains why, for given $k$, the asymmetric modes
feature smaller $N$: having smaller overlap with the region of $H>0$, they
need a smaller norm to compensate the shift of $k$ towards negative values
induced by the positive potential, see Eq. (\ref{stationaryNLSE}).

Note that the families of states displayed in Figs. \ref{bifurcation
diagrams}(a-c) comprise both $k<0$ and $k>0$, i.e., the leaky and trapped
modes alike. In particular, the SSB bifurcation occurs at $k>0$ in panels %
\ref{bifurcation diagrams}(a,b), and at $k<0$ in (c) (in the latter case,
the SSB sets in at $k=-0.100$, as shown for the same system in Fig. \ref%
{tails}). A noteworthy fact is that the system features the competition of
the two different mean-field phase transitions driven by the increase of $N$%
, i.e., the strength of the self-attraction: the transition from the
quasi-bound (leaky) state to the self-trapped one, which was previously
found in single-well elevated potentials \cite{Carr1,Carr2}, and the SSB in
the DWP structure. Thus, in the cases shown in Figs. \ref{bifurcation
diagrams}(a,b) the self-trapping transition happens first (at smaller $N$),
while in Fig. \ref{bifurcation diagrams}(c) the SSB takes place prior to the
onset of the self-trapping.

\begin{figure}[th]
\centering
\subfigure[] {\includegraphics[scale=0.3]{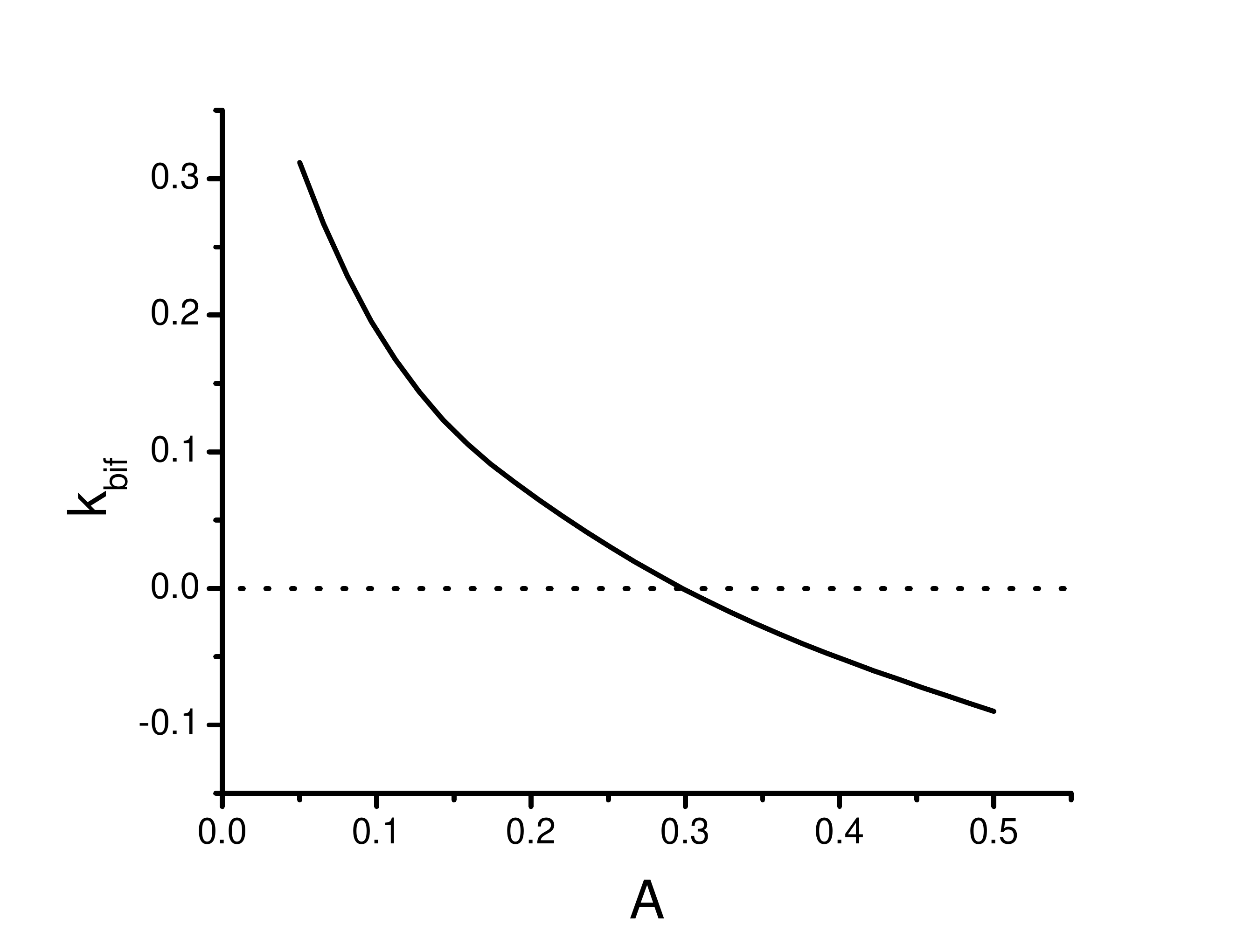} \label{bifurcation_k}}%
\nolinebreak
\subfigure[] {\includegraphics[scale=0.3]{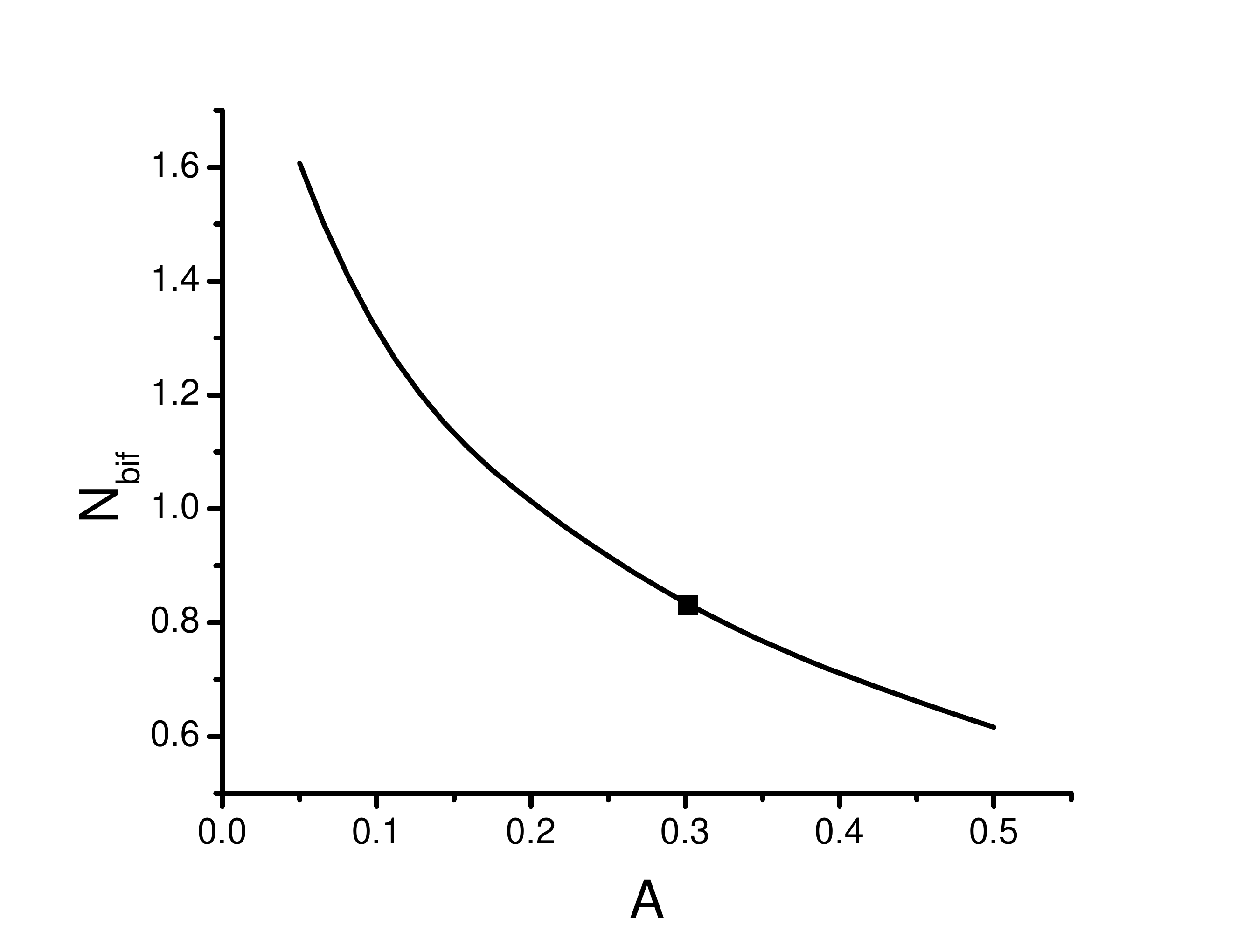}
\label{bifurcation_N}}
\caption{Values of propagation constant $k$ (a), and total norm $N$ (b), at
the SSB bifurcation point vs. the height of the inner barrier ($A$). The
horiziontal line, $k=0$, in (a), and the square symbol in (b) designate the
boundary between the trapped and leaky modes.}
\label{bifurcation point}
\end{figure}

The values of the propagation constant and norm at the SSB bifurcation point
are shown, as a function of the inner-barrier's height $A$ [see Eq. (\ref%
{potential})], in Fig. \ref{bifurcation point}. In panel (\ref{bifurcation_k}%
), the boundary between the SSB\ occurring with the delocalized and trapped
modes ($k_{\mathrm{bif}}<0$ and $k_{\mathrm{bif}}>0$, respectively) is
located at $A\approx 0.30$, the same value corresponding to the boundary
designated by the square symbol in panel \ref{bifurcation_N}). That is, the
SSB happens first (at smaller $N$) at $A>0.30$, while the transition to the
self-trapping precedes the SSB at $A<0.30$.

The fact that the SSB happens with the trapped and leaky modes,
respectively, at small and large $A$, as seen in Fig. \ref{bifurcation point}%
, is easy to explain: small $A$ implies strong linear coupling between the
wave functions in the two barely separated wells, hence very large $N$ is
required to induce the SSB, being far greater than the value of $N$ needed
for the onset of the self-trapping, which is determined by the fixed
parameters of the outer barrier in the DWP structure (\ref{potential}). On
the contrary, large $A$ implies weak linear coupling between the strongly
separated wells, hence the respective strength of the nonlinearity (measured
by $N$), required for the SSB, is much smaller than the value necessary for
the commencement of the self-trapping. These arguments clearly suggest that
the same sequence of the SSB and self-trapping phase transitions should take
place in generic DWP structures.

These arguments can be cast in a more definite form, if the central barrier
in Eq. (\ref{potential}) is approximated by $H_{\mathrm{central}}(x)=A\delta
(x)$, and the outer barriers are made impenetrable, similar to what is shown
in Fig. \ref{symmetric and asymmetric modes}(c). These conjectures replace
the present model by the one for an infinitely deep potential box split by
delta-functional barrier, which is the simplest model of the SSB \cite%
{NaturePhot2,Chili}. In the limit of large $A$, the latter model predicts
the following value of the norm at the SSB bifurcation point,%
\begin{equation}
N_{\mathrm{bif}}\approx 8\pi ^{2}/\left( 3l^{2}A\right)   \label{analyt}
\end{equation}%
(see caption to Fig. 2 in Ref. \cite{NaturePhot2}), where $l/2$ is the
coordinate of the point at which the wave function vanishes (half-width of
the infinitely deep box). In particular, one can identify $l\simeq 9$ for $%
A=0.5$ in Fig. \ref{symmetric and asymmetric modes}(b), hence Eq. (\ref%
{analyt}) yields $N_{\mathrm{bif}}~\simeq 0.65$, while the respective
numerical \ value in Fig. \ref{bifurcation point}(b) is $N_{\mathrm{bif}%
}~\approx 0.62$, which implies a reasonable agreement for the present (not
really large) value of $A$.

\section{Evolution of unstable symmetric states}

\begin{figure}[th]
\centering
\subfigure[] {\includegraphics[scale=0.13]{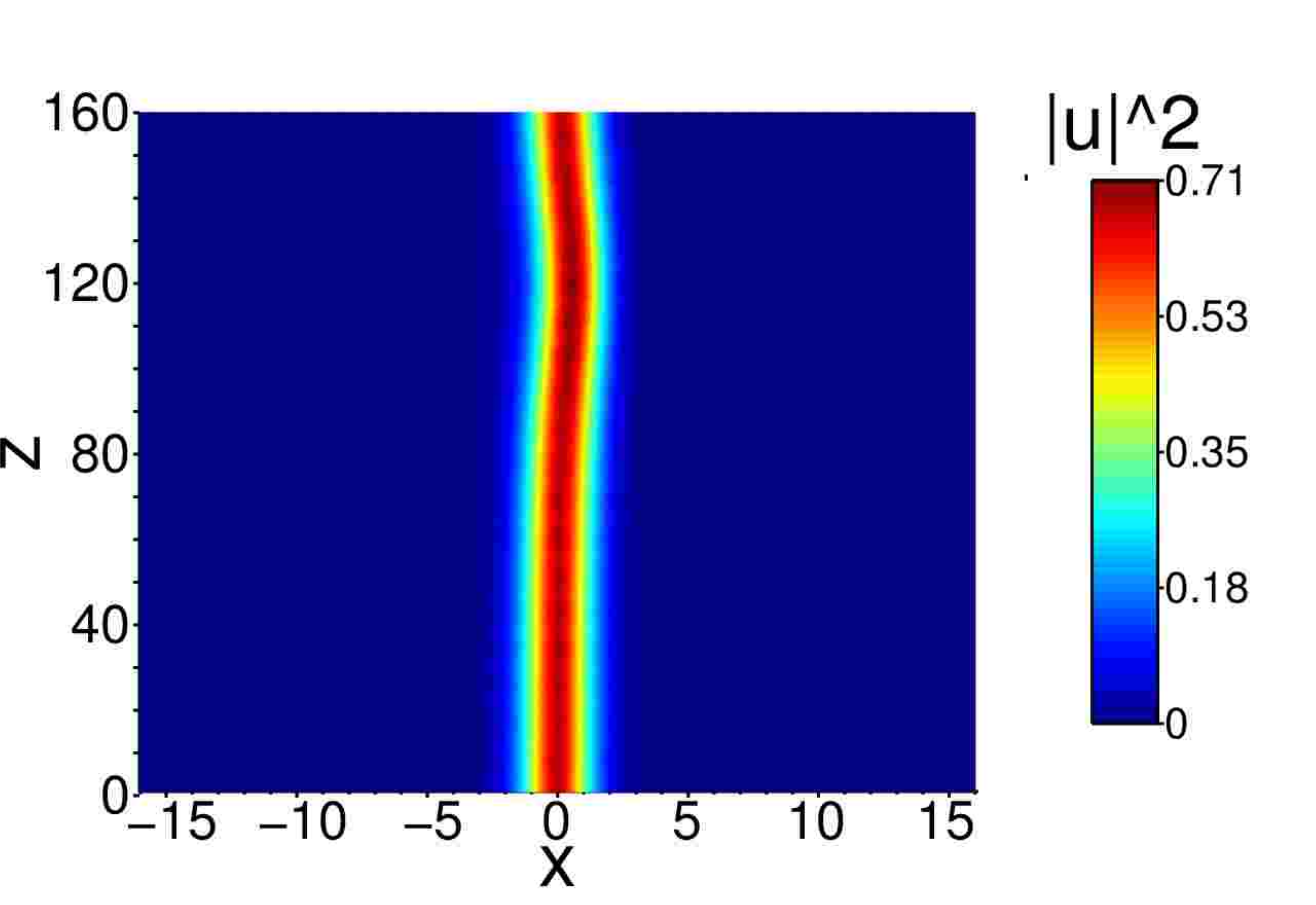}
\label{CH=0.05_k=0.314}}\nolinebreak \subfigure[]
{\includegraphics[scale=0.13]{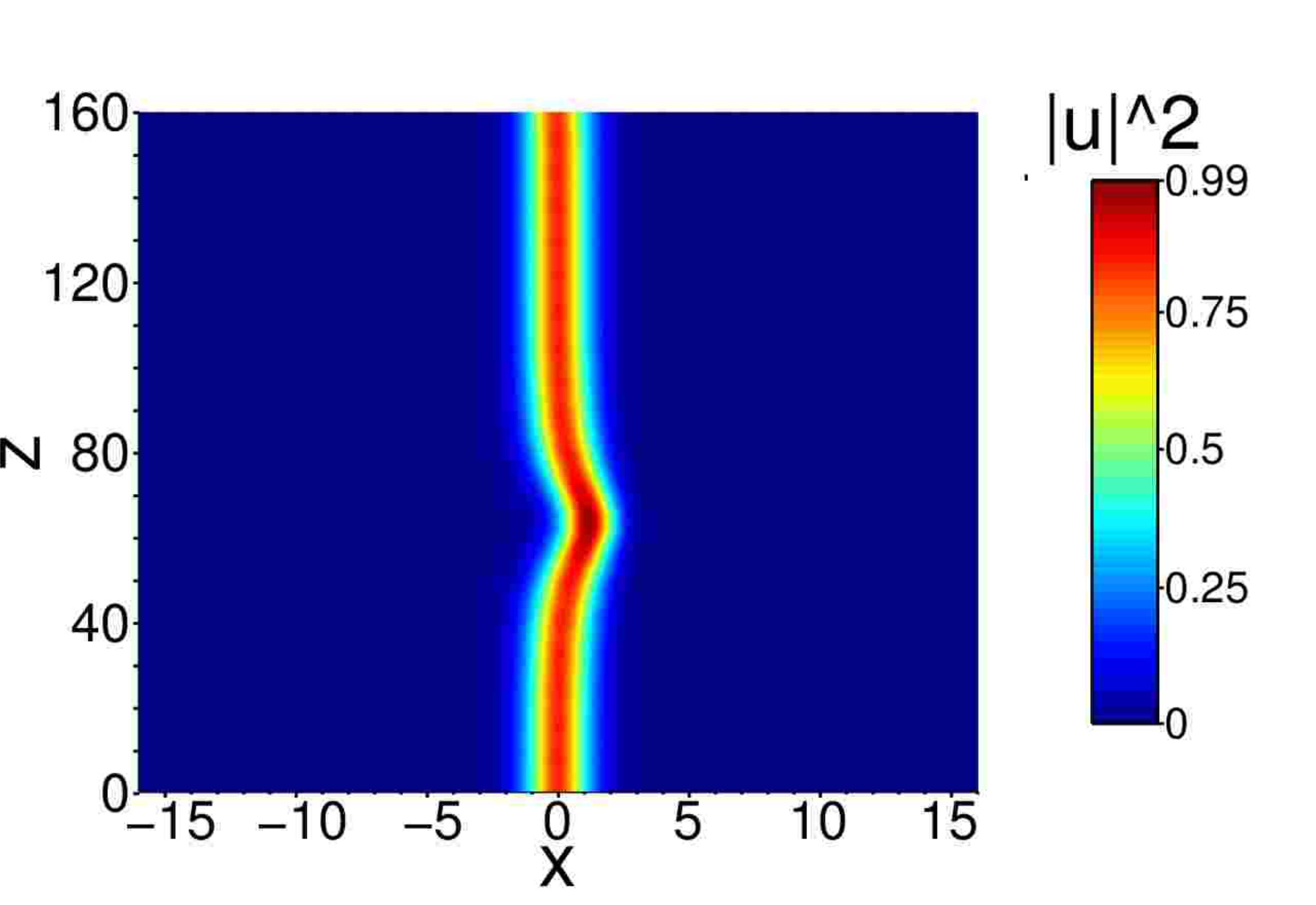}
\label{CH=0.05_k=0.40}}\nolinebreak \subfigure[]
{\includegraphics[scale=0.13]{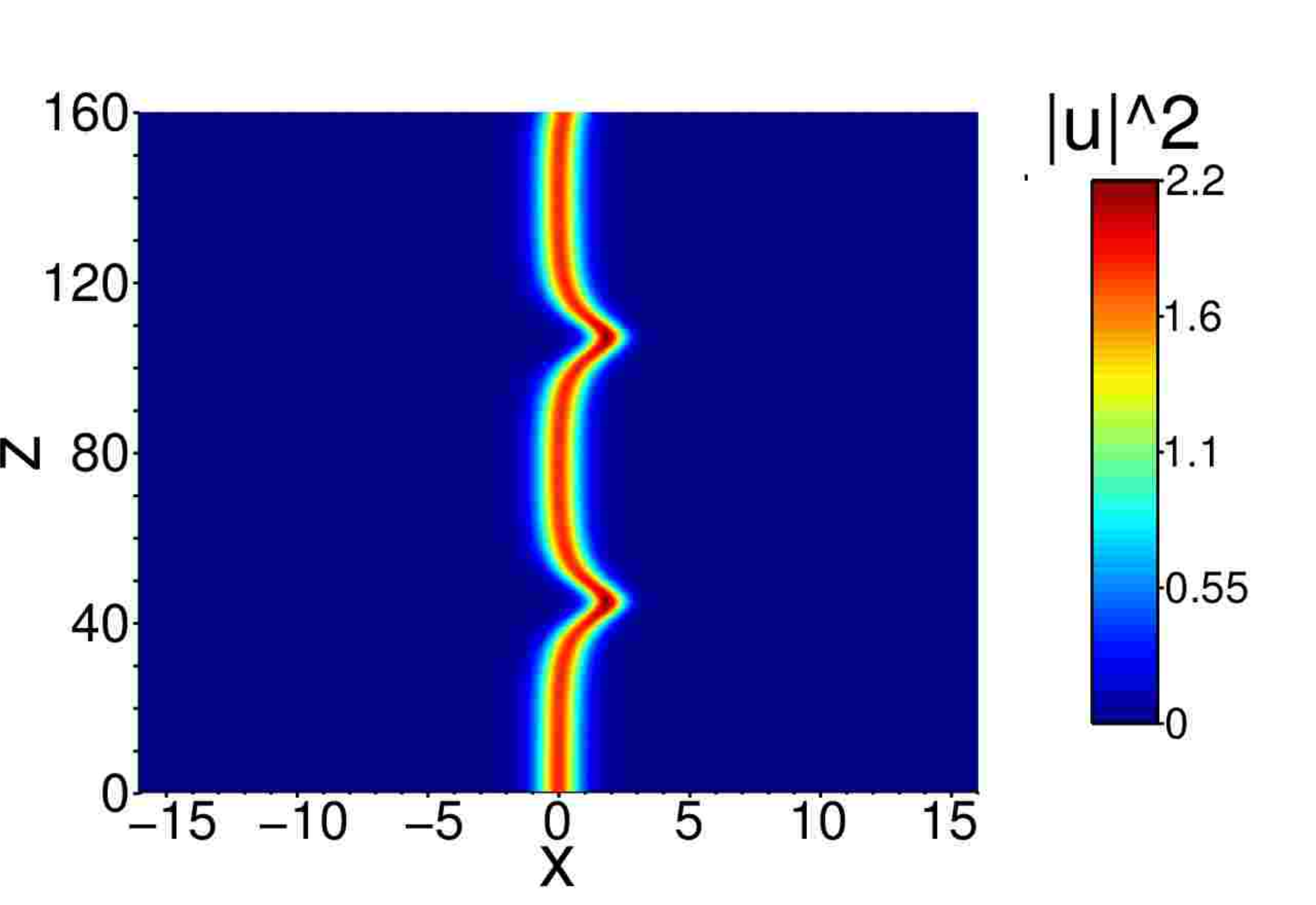}
\label{CH=0.05_k=0.90}}
\subfigure[] {\includegraphics[scale=0.13]{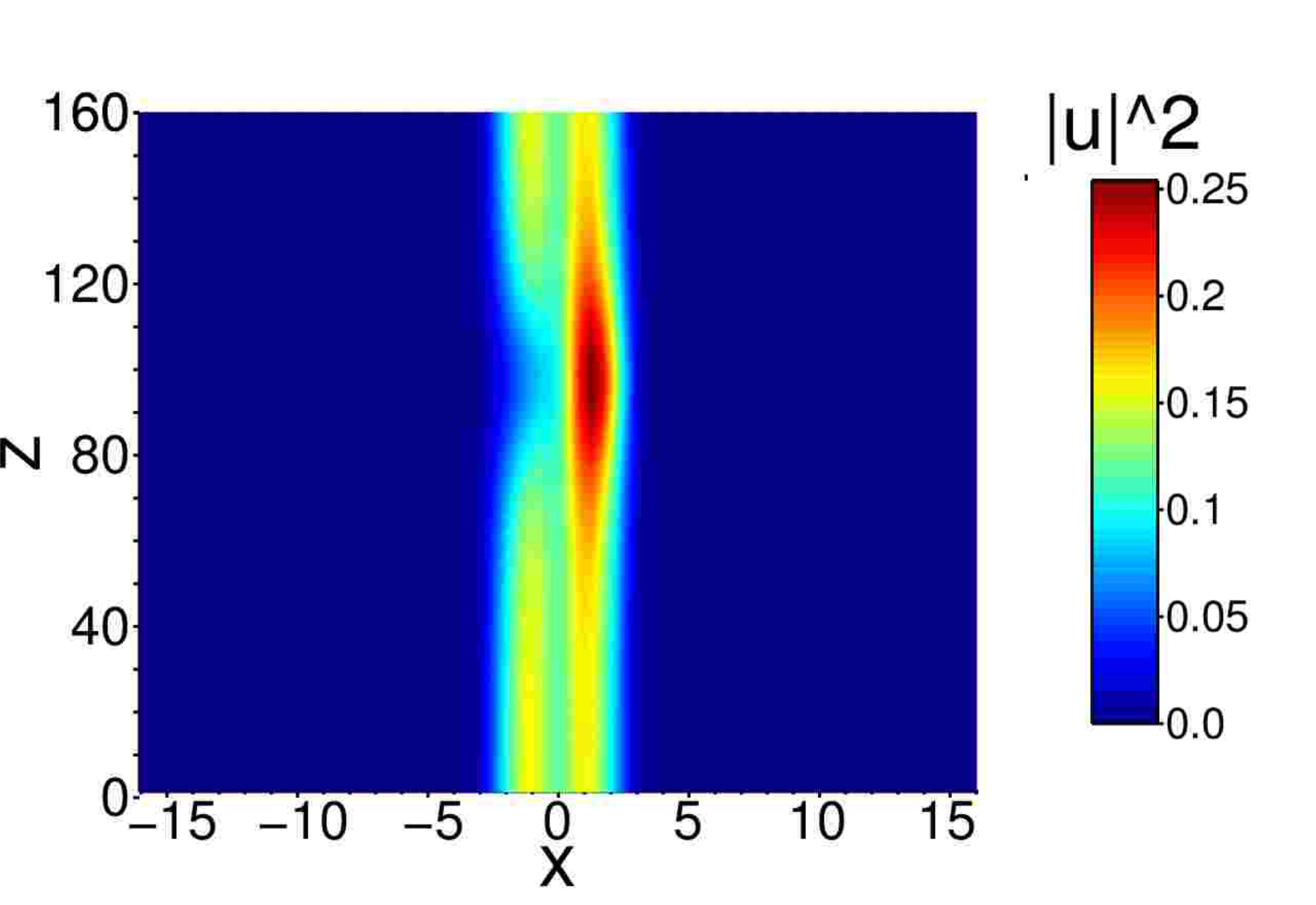}
\label{CH=0.5_k=-0.08}}\nolinebreak \subfigure[]
{\includegraphics[scale=0.13]{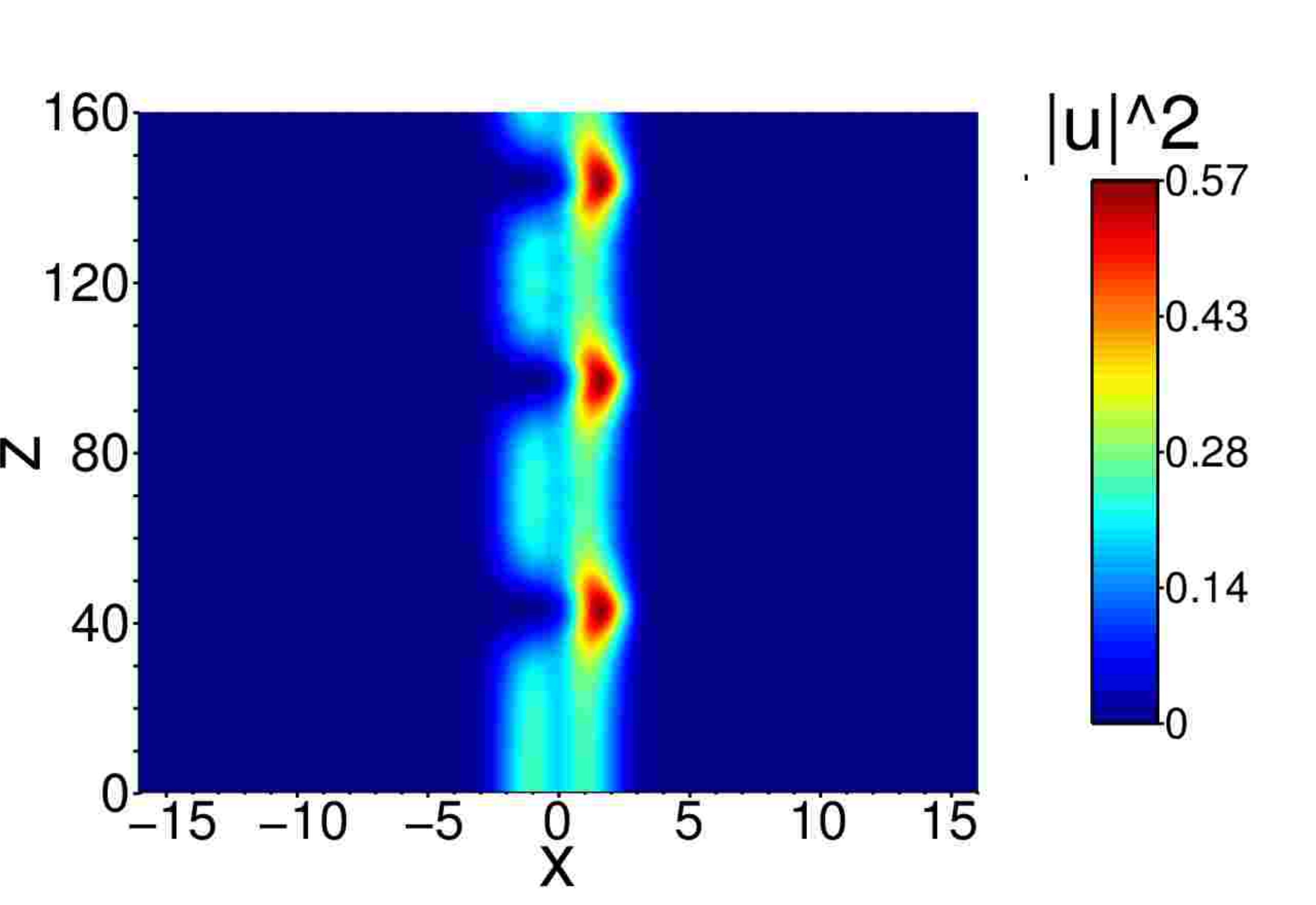}
\label{CH=0.5_k=-0.01}}\nolinebreak \subfigure[]
{\includegraphics[scale=0.13]{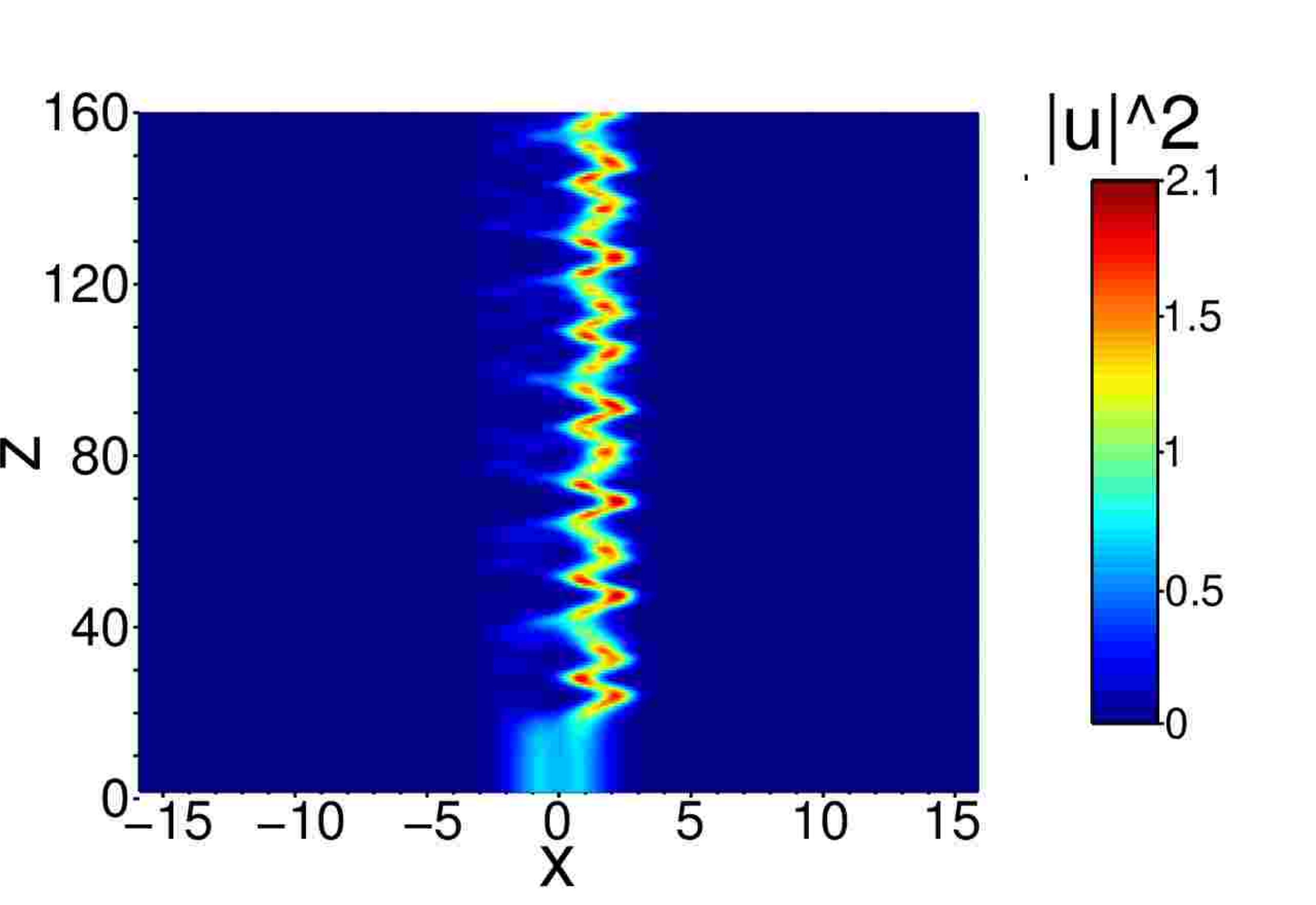}
\label{CH=0.5_k=0.30}}
\caption{(Color online) The evolution of unstable symmetric states is
displayed in panels (a-c) for potential (\protect\ref{potential}) with $%
A=0.05$, and in panels (d)-(f) for $A=0.5$, the corresponding SSB
bifurcations taking place at positive $k_{\mathrm{bif}}$, \textit{viz}., $k_{%
\mathrm{bif}}\approx 0.313$ (a-c), and negative $k_{\mathrm{bif}}$, \textit{%
viz}., $k_{\mathrm{bif}}\approx -0.100$ (d-f), respectively: (a) $k=0.314$;
(b) $k=0.40;$ (c) $k=0.90$; (d) $k=-0.08$; (e) $k=-0.01$; (f) $k=0.30$ [$k>0$
in (f) is chosen to display unstable evolution which takes place far from
the bifurcation point]. The instability gets stronger with the increae of $k$%
, i.e., moving from the left panels to the right ones.}
\label{unstable evolution}
\end{figure}

Conclusions concerning the stability and instability of the symmetric and
asymmetric modes, presented in Fig. \ref{bifurcation diagrams}, were
verified, in addition to the computation of eigenvalues for small
perturbations, by direct simulations, performed by dint of the
finite-difference algorithm. The instability development of unstable
symmetric states was catalyzed by adding small initial symmetry-breaking
perturbations to them. This was done for the unstable symmetric states with
both $k>0$ and for $k<0$, i.e., self-trapped and leaky ones. Typical
examples are displayed in Fig. \ref{unstable evolution}, in which the
unstable symmetric modes feature single- and double (split)-peak shapes at
small and large values of the inner-barrier's height, $A=0.05$ and $A=0.5$,
respectively (in the top and bottom rows of the figure).

As mentioned above, the instability of symmetric states is accounted for by
pure imaginary eigenfrequencies, hence the originally developing instability
is not oscillatory. However, the nonlinearity makes the unstable dynamics
oscillatory, as seen in Fig. \ref{unstable evolution}. In other words, the
unstable symmetric modes spontaneously develop bosonic Josephson
oscillations. Close to the instability onset, the effective oscillation
period is very large (as it diverges precisely at the onset point),
gradually decreasing deeper into the instability region. The dynamical
symmetry breaking induced by the weak and moderate instability is
incomplete, leading to periodic oscillations between the original symmetric
state and a new asymmetric one, as observed in Fig. \ref{unstable evolution}%
(a-e). Stronger instability causes complete symmetry breaking, replacing the
symmetric state by an irregularly vibrating asymmetric mode, as seen in Fig. %
\ref{unstable evolution}(f).

It is relevant to note the difference between the oscillatory regimes
generated by the instability of self-trapped and leaky modes. Indeed, while
Fig. \ref{unstable evolution}(c) demonstrates that the shape of the
oscillating mode is sharp in the former case,\ the shape is fuzzy in Fig. %
\ref{unstable evolution}(e) because it involves an essential radiation
component in the case when the underlying unstable mode is a leaky one.

\section{Collisions of free solitons with the quasi-double-well potential
structure}

\label{sec:CollisionResults}

In addition to the analysis of the stationary states performed above, it is
also relevant to consider collisions of free solitons with the elevated DWP
structure. For this purpose, the initial soliton was created as the tilted
(moving) version of the static one given by Eqs. (\ref{psi}) and (\ref%
{soliton}),%
\begin{equation}
\psi \left( x,z\right) =\sqrt{2k}\exp \left( i\left( k-c^{2}/2\right)
z+icx\right) \text{\textrm{sech}}\left( \sqrt{2k}\left( \left(
x-x_{0}\right) -cz\right) \right) ,  \label{tilt}
\end{equation}%
where $c$ is a real tilt (velocity), and $x_{0}$ is the initial position of
the soliton, chosen far enough from the localized potential structure (we
here take $x_{0}=-15$). Generic findings are produced here for incident
solitons with $k=3,$ the corresponding norm being
\begin{equation}
N=2\sqrt{2k}\approx 4.90,  \label{N}
\end{equation}%
according to Eq. (\ref{sol}), other values of $N$ giving similar results.

\begin{figure}[tbp]
\includegraphics[width=3.2in]{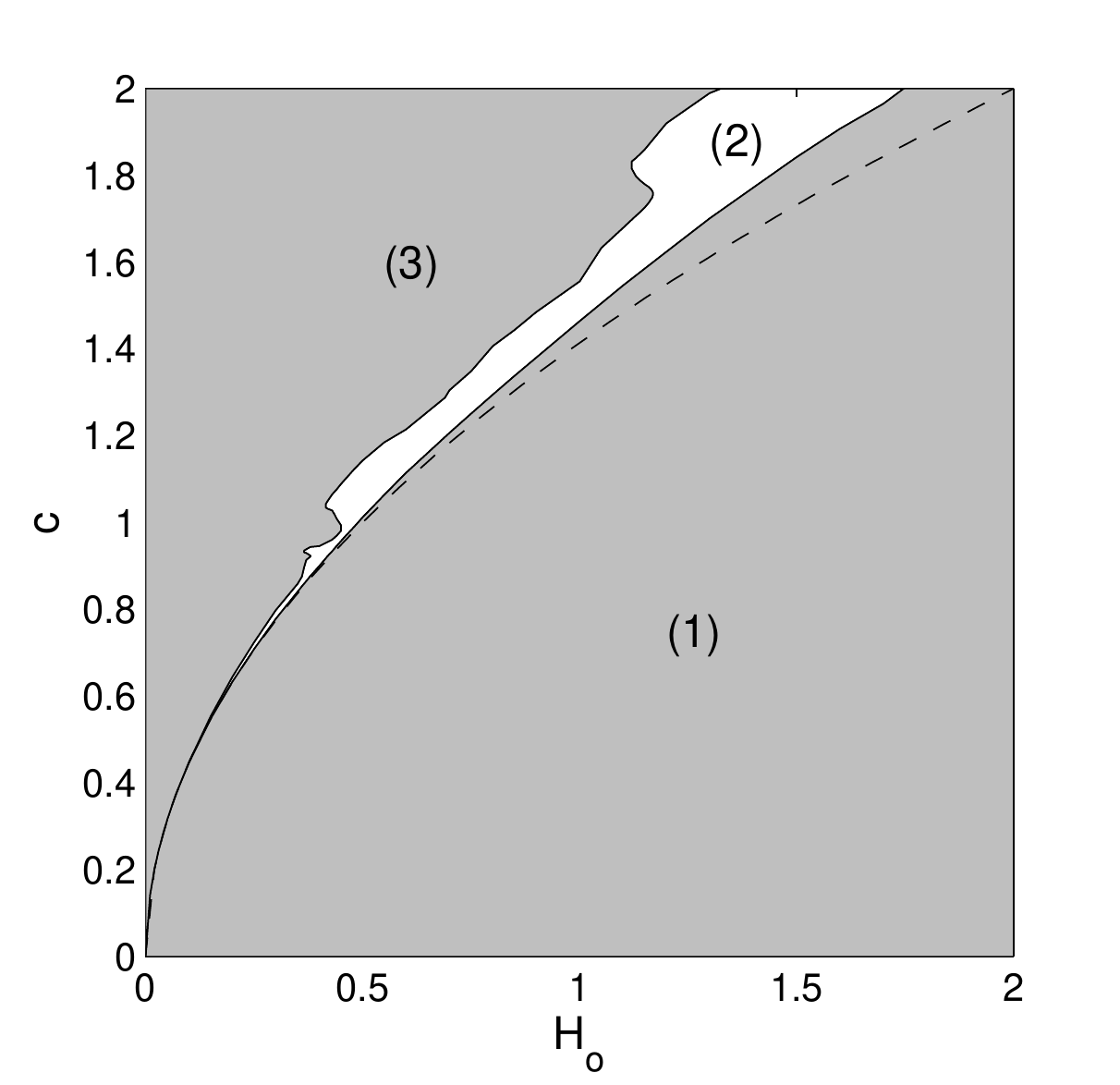}%
\caption{The chart of outcomes of collisions of free solitons, launched with
tilt $c$ [see Eq. (\protect\ref{tilt})], with potential structure (\protect
\ref{P2}), in the plane of $\left( H_{0},c\right) $. The norm of the
incident soliton is fixed as per Eq. (\protect\ref{N}). In this figure, the
width of the outer potential barriers is $W_{0}=4$. In region (1), the
soliton bounces back from the left outer barrier. Region (2) refers to
trapping the soliton inside the potential structure, where it performs
shuttle motion. In region (3), the incident soliton passes the structure.
The dashed line is the analytical prediction produced by Eq. (\protect\ref%
{c_cr}).}
\label{ThetaVsHeightMapping}
\end{figure}

\begin{figure}[tbp]
\subfigure[]{%
\includegraphics[width=2.3in]{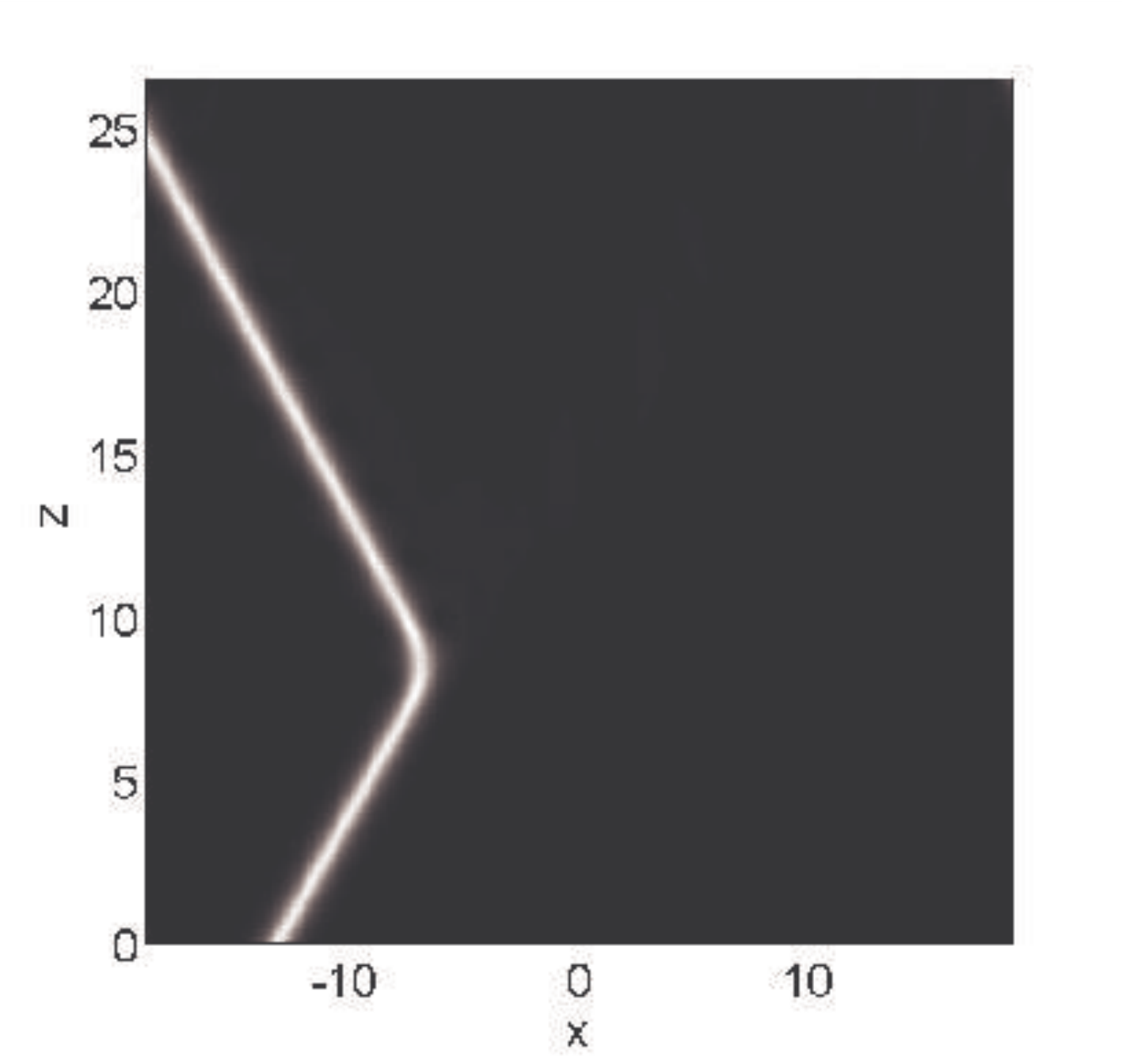}%
\label{FundCollision1}} \subfigure[]{%
\includegraphics[width=2.3in]{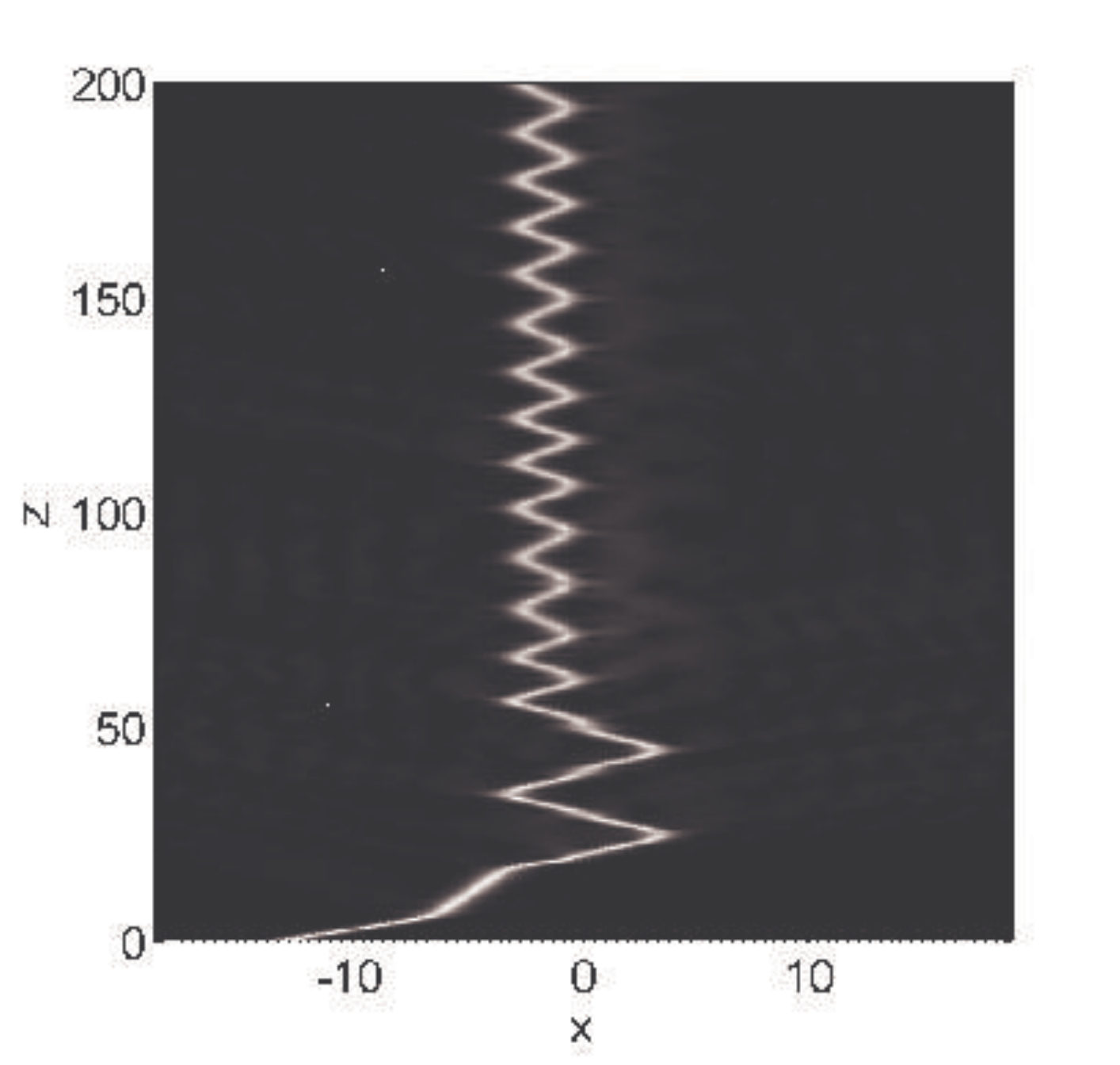}%
\label{FundCollision2}} \subfigure[]{%
\includegraphics[width=2.3in]{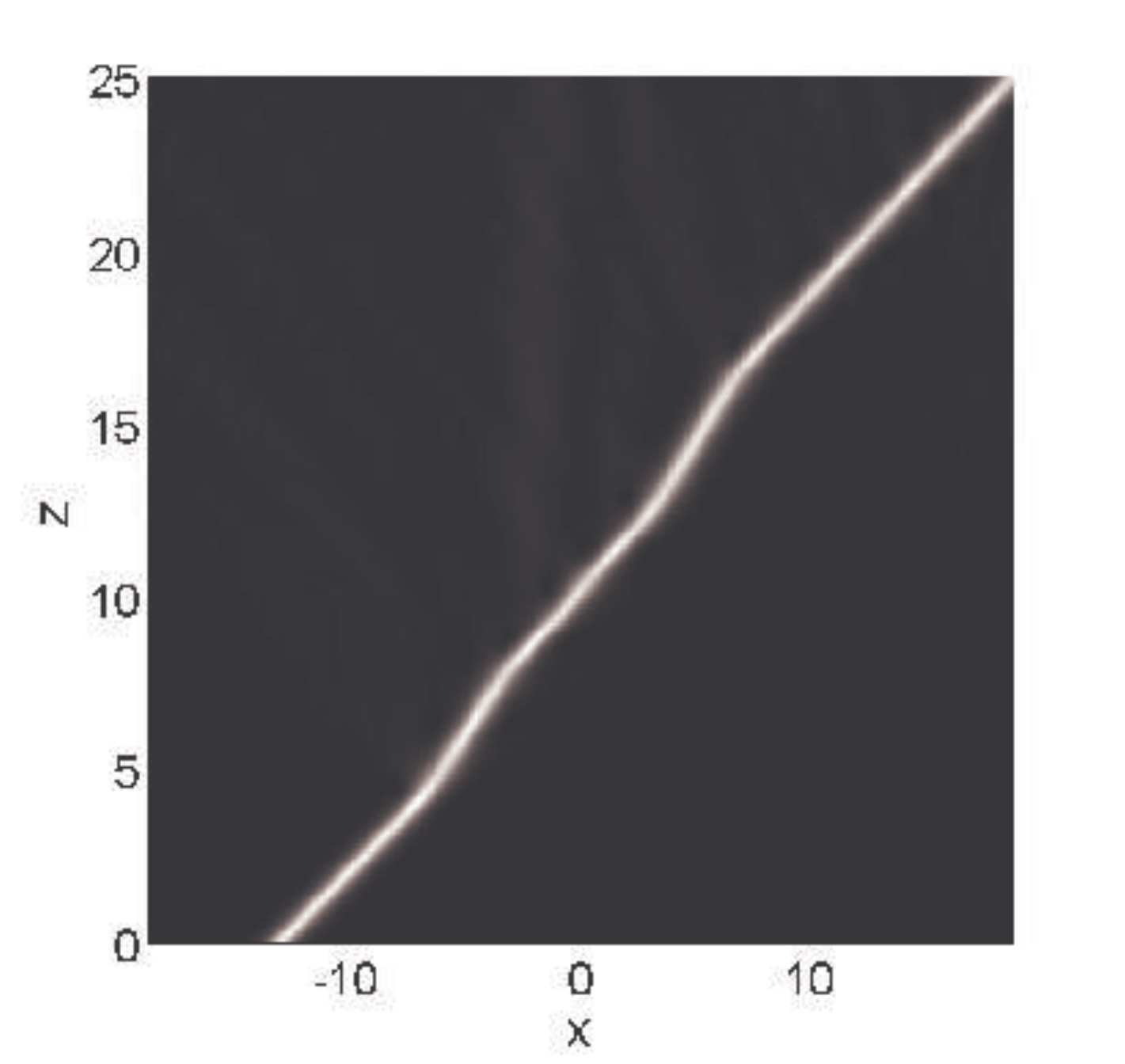}%
\label{FundCollision3}} \subfigure[]{%
\includegraphics[width=2.3in]{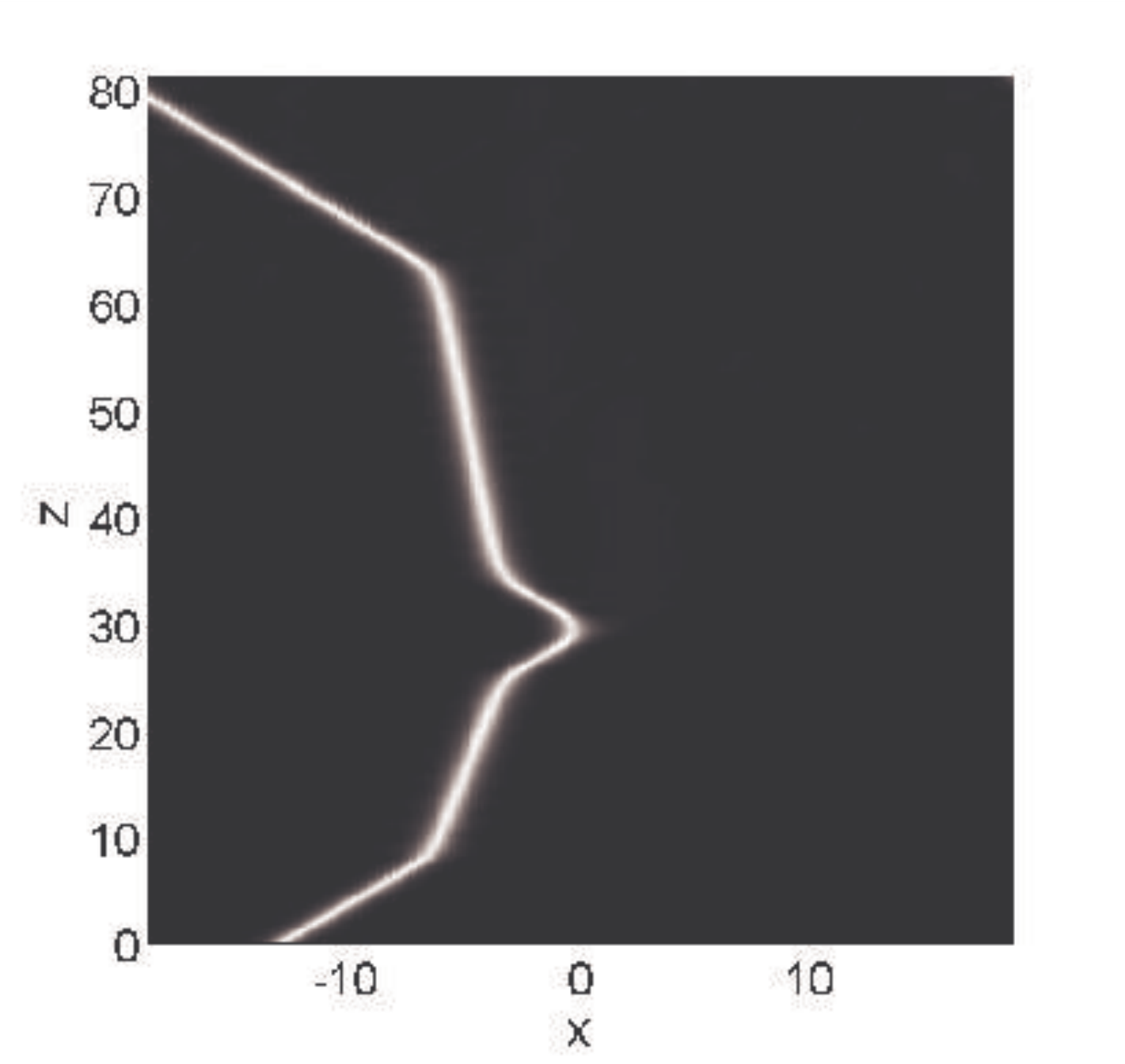}%
\label{FundCollision4}} \subfigure[]{%
\includegraphics[width=2.3in]{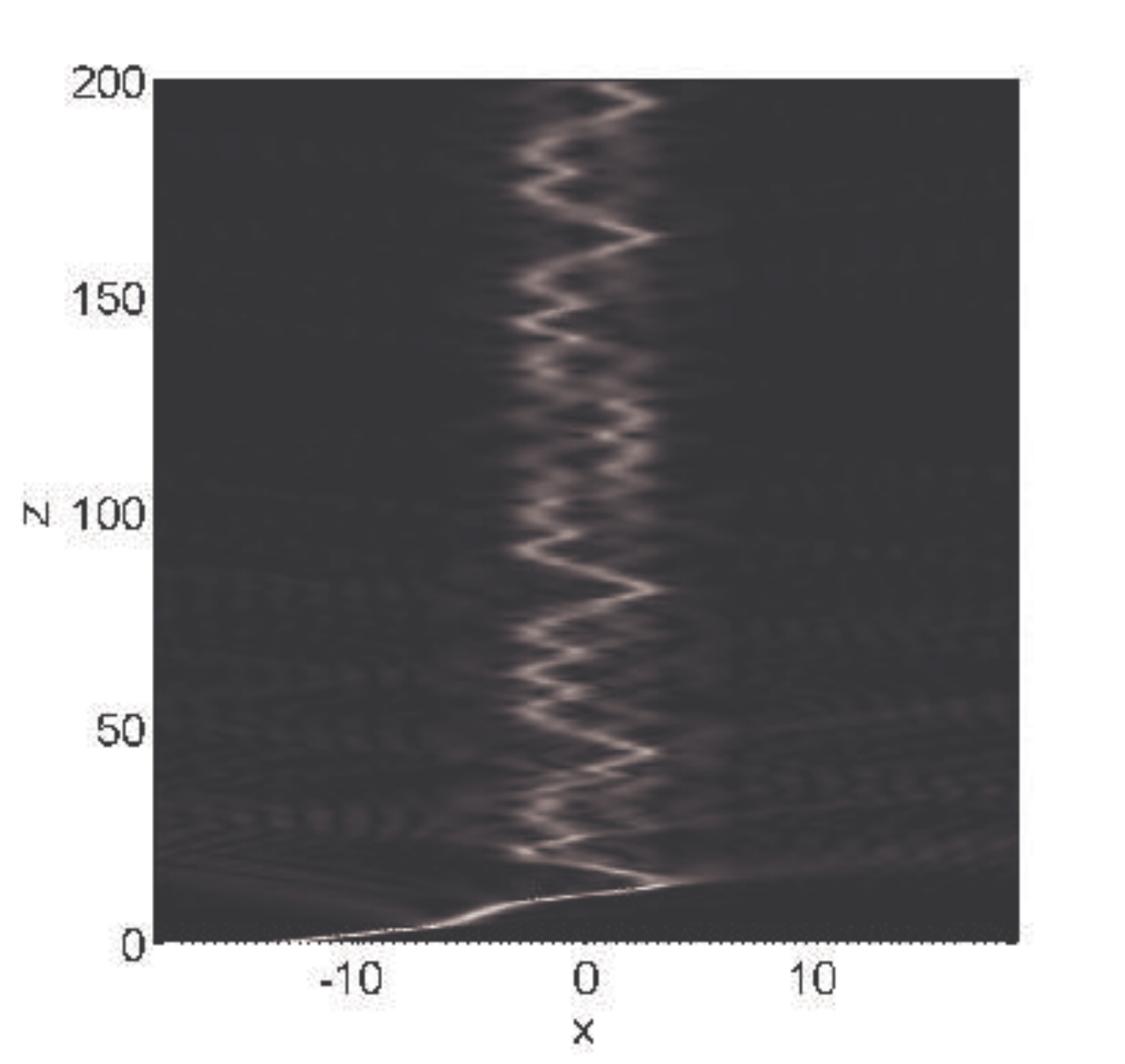}%
\label{FundCollision5}} \subfigure[]{%
\includegraphics[width=2.3in]{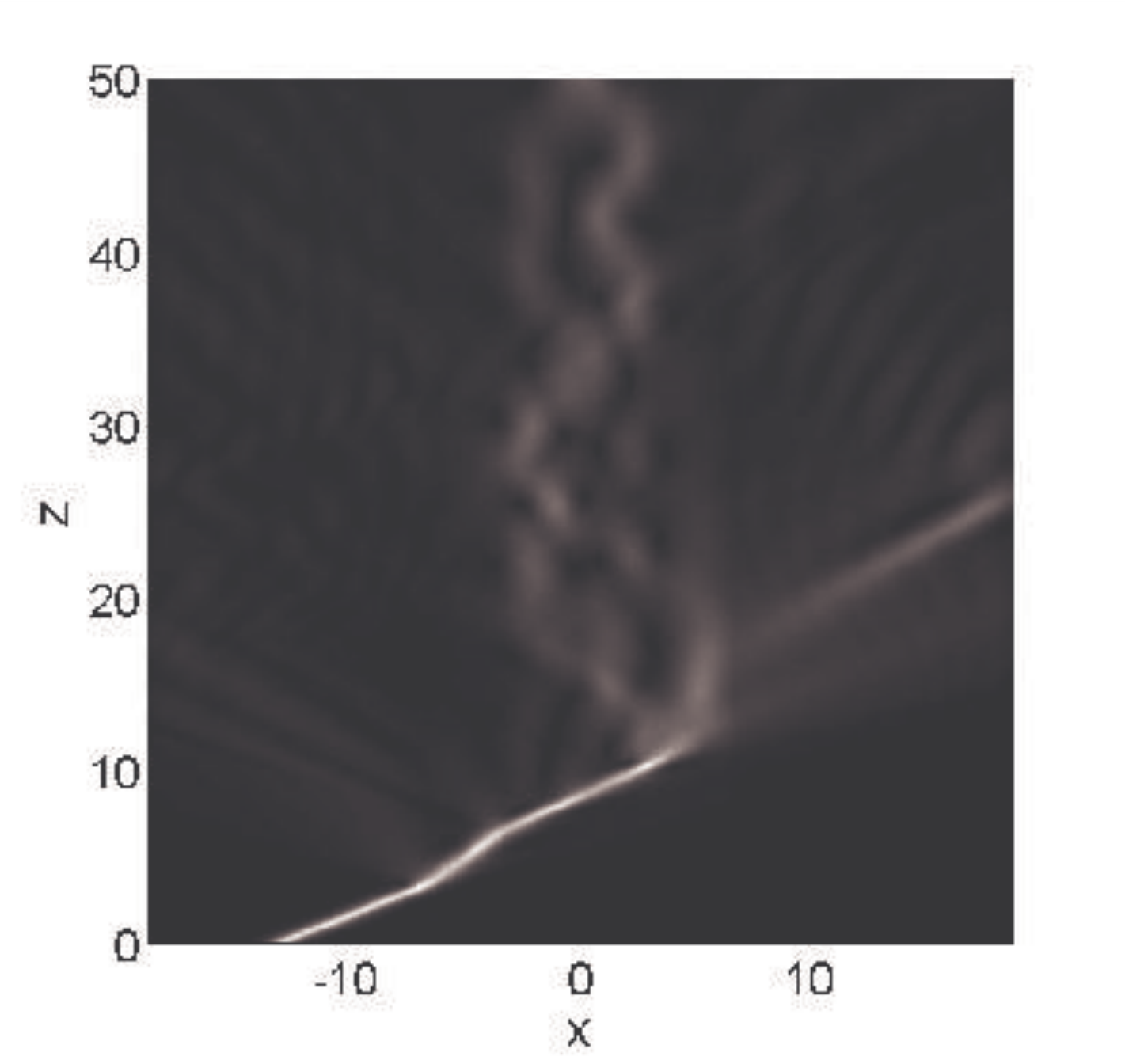}%
\label{FundCollision6}}
\caption{Generic examples of outcomes of collisions of the soliton with the
double-well potential structure (\protect\ref{P2}), for $N=4.90$ ($k=3$), $%
W_{0}=4$, and different heights of the outer initial barriers, $H_{0}$, and
different tilts $c$ of the incident soliton: (a) $[H_{0}=0.6,c=0.8]$, (b) $%
[H_{0}=0.6,c=1.15]$, (c) $[H_{0}=0.6,c=1.5]$, (d) $[H_{0}=0.3,c=0.8]$, (e) $%
[H_{0}=1.3,c=1.8]$, (f) $[H_{0}=1.2,c=1.92]$, see further explanations in
the text.}
\label{FundCollisions}
\end{figure}

Figure \ref{ThetaVsHeightMapping} presents a parameter chart for three
different outcomes of the collisions, produced by varying the height of the
outer barriers, $H_{0}$ in Eq. (\ref{P2}), and tilt $c$ in Eq. (\ref{tilt}).
In the region designated as (1), i.e., for $c$ small and/or $H_{0}$ large
enough, the incident soliton bounces back the left outer barrier, as shown
in Fig. \ref{FundCollision1} for $H_{0}=0.6$ and $c=0.8$. At larger $c$, in
a relatively narrow region (2) of Fig. \ref{ThetaVsHeightMapping}, the
soliton gets captured inside the potential structure, which resembles the
previously explored possibility of capturing an incident Bragg-grating
soliton by a cavity formed by two locally repulsive defects \cite{cavity}.
Similar to that setting, the trapped soliton performs shuttle motion between
the outer and inner potential barriers, as shown in Fig. \ref{FundCollision2}%
, for $H_{0}=0.6$ and $c=1.15$. The shuttle dynamical regime is an essential
addition to the stationary states revealed by the analysis presented above.
At still larger $c$, but yet staying within the boundaries of the narrow
region (2), the shuttle motion of the trapped soliton becomes irregular, see
Fig. \ref{FundCollision5} for $H_{0}=1.3$ and $c=1.8$.

In region (3), the initial tilt, $c$, is large enough for the soliton to
pass the potential structure, see an example in Fig. \ref{FundCollision3}
for $H_{0}=0.6$ and $c=1.5$. Another option admitted by this scenario is
shown in Fig. \ref{FundCollision4}, where the soliton passes the left outer
barrier, bounces back from the inner one, and escapes in the reverse
direction, passing the left outer barrier again. Naturally, this collision
pattern is common for lower $H_{0}$, when the outer barriers are \emph{lower}
than the inner one, which plays the role of a strong ``bouncer".
Furthermore, at $c>1.85$, the incident soliton splits into two fragments,
one escaping and the other one staying in a chaotically evolving trapped
state, see an example in Fig. \ref{FundCollision6}, for $H_{0}=1.2$ and $%
c=1.92$.

Collision scenarios were also explored by varying width $W_{0}$ of the outer
barriers, while keeping their height constant, $H_{0}=1$, as well as
characteristics of the inner barrier and the distance between the barriers,
see Eq. (\ref{P2}). The respective results, for the same incident soliton as
used above ($k=3$, $N=4.90$), are summarized in Fig. \ref%
{ThetaVsWidthMapping}, where regions (1)-(3) have the same meaning as their
counterparts in Fig. \ref{ThetaVsHeightMapping}.

In the latter case, the results may be classified into three outcomes of the
collision, depending on $W_{0}$. The first outcome occurs at $0<W_{0}<0.8$.
It is characterized by a rapidly growing region of the shuttle motion of the
trapped soliton [region (2)], and \emph{sharp transitions} between the three
evolution scenarios, $(1)\leftrightarrow (2)$, $(2)\leftrightarrow (3)$. The
second outcome, which was observed in the range of $0.8<W_{0}<1.8$ (in this
case, $W_{0}$ is, roughly, close to the width of the incident soliton), is
distinguished by \emph{gradual transitions} between the scenarios. That
means that, for certain values of $c$, the soliton does not fully bounce
from the barrier, nor penetrates it, but rather splits into two segments,
one of which escapes, while the other remains trapped. In this case, both
the upper and lower boundaries of region (2) represent tilts at which the
soliton is split into equal fragments. An example of such an outcome is
shown in Fig. \ref{SplitSolitonCollision}, for $W_{0}=1.2$ and $c=1.66$. The
third outcome, which is observed at $W_{0}>1.8$, is distinguished by the
fact that the variation of $W_{0}$ almost does not affect the soliton's
motion. In contrast to the second outcome, and similar to the first one, the
respective transitions between the regions are sharp.

\begin{figure}[tbp]
\includegraphics[width=3.2in]{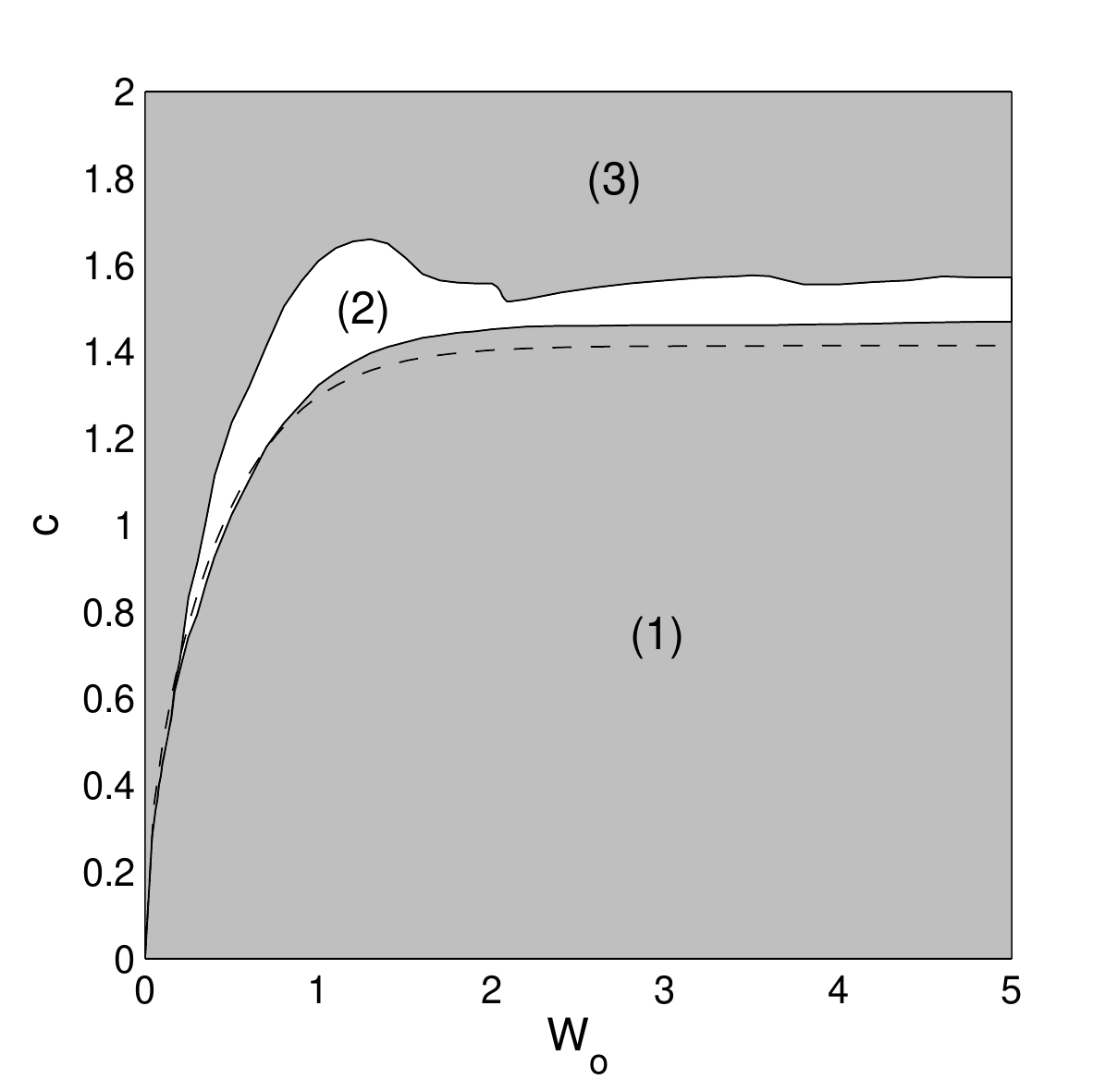}
\caption{The same as in Fig. \protect\ref{ThetaVsHeightMapping}, but varying
the width of the outer potential barriers, $W_{0}$, while their height is
fixed, $H_{0}=1$. The dashed curve shows the analytical prediction for the
boundary of area (1) given by Eq. (\protect\ref{parabola}). }
\label{ThetaVsWidthMapping}
\end{figure}

\begin{figure}[tbp]
\includegraphics[width=3.2in]{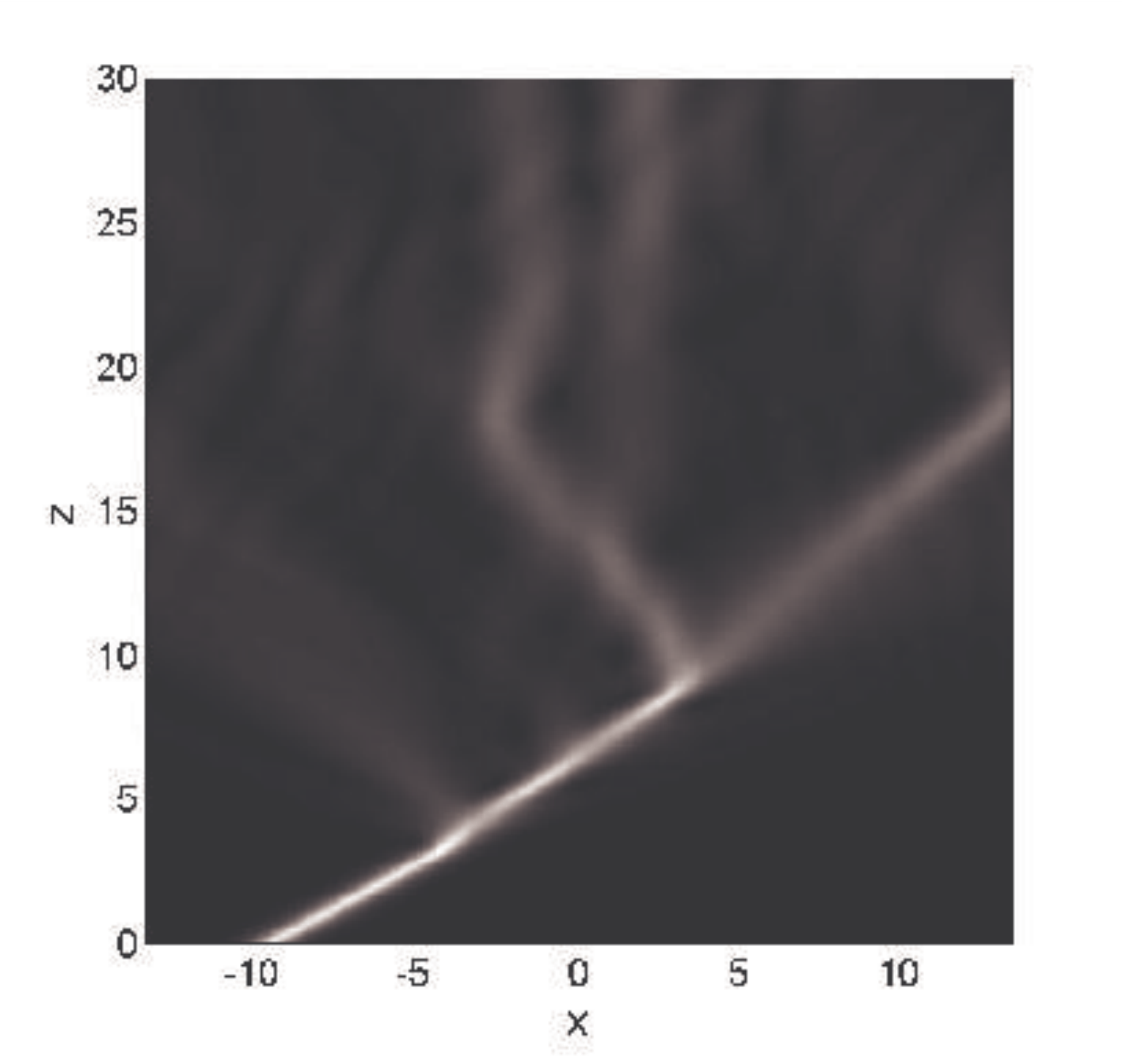}
\caption{Outcomes of the collisions, for $W_{0}=1.2$ and $c=1.66$. In this
borderline example, the soliton hits the right (second) outer potential
barrier and splits into two fragments, one escaping and the other one
staying trapped in the potential structure.}
\label{SplitSolitonCollision}
\end{figure}

It is easy to explain the parabolic boundary of region (1) in Fig. \ref%
{ThetaVsHeightMapping}, as well as the boundary of the same region in Fig. %
\ref{ThetaVsWidthMapping}, using the perturbation theory for NLS solitons,
which treats them as quasi-particles with mass $N$ \cite{RMP}. Indeed, the
kinetic energy of the soliton is $E_{\mathrm{kin}}=(N/2)c^{2}$, while the
height of the outer potential barrier for the quasi-particle, $E_{0}$, can
be obtained from the third term in expression (\ref{H}), assuming that the
soliton's center is located at the midpoint of the potential barrier:%
\begin{equation}
E_{0}=H_{0}\int_{-W_{0}/2}^{+W_{0}/2}u_{\mathrm{sol}}^{2}\left( x^{\prime
}\right) dx^{\prime }=H_{0}N\tanh \left( \frac{1}{4}W_{0}N\right) ,
\label{E0}
\end{equation}%
where $u_{\mathrm{sol}}$ is the solitons's profile (\ref{soliton}), $%
x^{\prime }\equiv x-\left( 3+W_{0}/2\right) $ [see Eq. (\ref{P2})], and the
result is expressed in terms of the soliton's norm, as per Eq. (\ref{sol}).
Next, equating $E_{\mathrm{kin}}$ to $E_{0}$ predicts that the boundary
between the rebound and passage corresponds to%
\begin{equation}
c_{\mathrm{cr}}=\sqrt{2H_{0}\tanh \left( \frac{1}{4}W_{0}N\right) }.
\label{parabola}
\end{equation}%
In particular, the respective prediction for dependence $c_{\mathrm{cr}%
}(H_{0})$ corresponding to the case displayed in Fig. \ref%
{ThetaVsHeightMapping}, with $W_{0}=4$ and $N$ fixed as per Eq. (\ref{N}),
simplifies to%
\begin{equation}
c_{\mathrm{cr}}=\sqrt{2H_{0}}.  \label{c_cr}
\end{equation}%
Figure \ref{ThetaVsHeightMapping} demonstrates that Eq. (\ref{c_cr})
predicts the parabolic boundary of region (1) quite accurately, a
discrepancy at large $H_{0}$ being explained by the fact that the collision
with the tall barrier causes a deformation of the soliton's shape. Further,
the full analytical expressions (\ref{parabola}) predicts the boundary of
the same region in Fig. \ref{ThetaVsWidthMapping} well enough too.

As concerns the trapping regime in area (2) of Figs. \ref%
{ThetaVsHeightMapping} and \ref{ThetaVsWidthMapping}, it is explained by the
fact that, while slowly passing the left barrier, and then passing the inner
one, the soliton with the initial tilt slightly exceeding value (\ref%
{parabola}) suffers radiation losses due to its deceleration and
acceleration. The losses cause a drop in the kinetic energy below the level
necessary for clearing the right barrier. A feature which relates the
trapping and splitting to the leaky modes considered above is tunneling of
the radiation across the outer potential barriers, which can be seen, in
particular, in Figs. \ref{FundCollisions}(b,e,f).

\section{Conclusions}

We have extended the known possibility of the stabilization of leaky modes
in quasi-trapping potentials by means of the self-attractive nonlinearity.
Unlike the previously studied single-well potential, we have introduced the
spatially symmetric DWP (double-well potential) with the elevated floor,
embedded into the potential barrier. The setting can be implemented in
nonlinear optical waveguides and BEC. The new possibility offered by this
system is the competition of two phase transitions of the second kind,
described in the mean-field approximation: the onset of the self-trapping of
the leaky modes, and the SSB (spontaneous symmetry breaking) of both true
bound states and leaky modes, under the action of the self-attractive
nonlinearity. With the increase of the norm of the wave field (which
determines the nonlinearity strength), the former or latter transition
happens first if the central barrier of the DWP structure is, respectively,
low or tall. These conclusions are generic, as they do not depend on details
of the DWP structure. New effects are revealed by the consideration of the
SSB of the leaky modes: asymmetry of radiation tails, which are parts of
these modes, and the commensurability-incommensurability interplay between
the radiation wavelength and the total size of the system, into which the
DWP is embedded. Systematic results have been produced in the numerical
form, and their basic features were explained with the help of analytical
considerations. The simulations demonstrate that unstable symmetric modes
initiate Josephson oscillations. Collisions of freely moving solitons with
the DWP structure were studied in a systematic form too, revealing various
generic outcomes of the collisions, boundaries between which were explained
in the analytical form. In particular, in addition to the stationary states
with the unbroken and broken symmetry, the collisions reveal the dynamical
mode, in the form of a soliton which performs persistent shuttle motion in
the DWP structure.

Relevant possibilities for the extension of the analysis reported in this
work may be offered by two-component systems, as well as by a
two-dimensional generalization of the present setting. On the other hand,
the results produced by the competition of the mean-field phase transitions
suggest that it may be interesting too to consider quantum phase transitions
in a many-body bosonic state with attractive inter-particle interactions,
loaded into the DWP.

\section{Acknowledgment}

M.T. and B.A.M. acknowledge partial support from the National Science Center
of Poland in the framework of the HARMONIA Program, No. 2012/06/M/ST2/00479.
K.B.Z. acknowledges support from the National Science Center of Poland
through Project ETIUDA No. 2013/08/T/ST2/00627.

\end{document}